\documentclass[12pt]{iopart}

\usepackage[lofdepth,lotdepth]{subfig}
\usepackage{graphicx}

\usepackage{algorithm}
\usepackage{algpseudocode}
\usepackage{color}
\usepackage{cite}
\usepackage{amssymb}
\usepackage{gensymb}
\usepackage{hyperref}
\usepackage{multicol}

\newcommand{\R}[1]{{\color{black}#1}}

\newcommand{\half}{\frac{1}{2}}

\newcommand{\ten}[1]{\overline{\overline{#1}}}

\begin{document}

\title[A novel and fast electromagnetic and electrothermal software ...]{A novel and fast electromagnetic and electrothermal software for quench analysis of high field magnets\footnote{This is the Accepted Manuscript version of an article accepted for publication in Superconductor Science and Technology. IOP Publishing Ltd is not responsible for any errors or omissions in this version of the manuscript or any version derived from it. The Version of Record is available online at \url{https://doi.org/10.1088/1361-6668/ad68d3}. This manuscript is hereby available under license CC BY-NC-ND.}}

\author{Anang~Dadhich$^1$,
				Philippe~Fazilleau$^2$ and
				Enric~Pardo$^1$$^{+}$}
				
\address{$^1$Institute of Electrical Engineering, Slovak Academy of Sciences, Bratislava, Slovakia\\
$^2$CEA-Saclay, Paris, France\\
$^+$Author to whom correspondence should be addressed (enric.pardo@savba.sk)}

\begin{abstract}
High-field {superconducting} REBCO magnets contain several coils with many turns. For these magnets, electro-thermal quench is an issue that magnet designers need to take into account. Thus, there is a need for a fast and accurate software to numerically model the overall performance of full-scale magnets. High temperature superconductors can be modeled using different techniques for electro-magnetic and thermal (finite element method) analysis. However, it takes a lot of time to model the electro-magnetic and electro-thermal behavior of superconductors simultaneously, especially for non-insulated or metal-insulated coils. {{In addition, most} of the available methods ignore screening currents, which are an important feature of {REBCO magnets}}. We have developed a novel software programmed in C++, which performs coupled electro-magnetic and electro-thermal analysis using variational methods based on Minimum Electro-Magnetic Entropy Production (MEMEP) and Finite Difference, respectively. The developed {software, which takes screening currents into account,} is applied to axi-symmetric fullscale magnets of more than {32 T} field strength under {the} {SuperEMFL project} for {thermal quench reliability during standard operation}. We show that the magnets incorporating non-insulated coils are more reliable against quench than the metal insulated coils. Also, realistic cooling conditions at boundaries is essential for such simulations. The model developed can be used for a quick and complete electro-magnetic and electro-thermal analysis of superconducting high field magnets. 
\end{abstract}

\section{Introduction}

High field superconducting magnets have many present and potential future applications, such as medical Magnetic Resonance Imaging (MRI) machines \cite{iwasabook2009, lvovskySuST2013, JimenoNMR2022}, particle accelerators like Large Hadron Collider for research \cite{botturaFP2022, rossiReviews2012, tollestrupReviews2008}, fusion reactors for energy applications \cite{yanagiNF2015,ITOFE2018, coataneaIEEE2015}, and high field scientific experiments{, which} require tens of Teslas of magnetic field strength \cite{hahn2019Nat, wangSuST2021, awajiIEEE2013, superemfl2023,  dadhichThesis2021}. Though these magnets are able to generate high field (above 30 T), they are also prone to limiting effects {like} electrothermal quench{. This kind of quench} can occur when the current in superconductors goes over the critical current ($I_c$) of a certain turn, which causes quick temperature rise. This sudden sharp increase in temperature can result in thermal runaway, or even burning of the HTS (High Temperature Superconducting) coils or the whole high field magnets. {In addition, HTS windings are inserted at the bore of Low Temperature Superconducting  (LTS) magnets (or outserts) that generate a background magnetic field to the HTS winding.} Thus, the study of electro-magneto-thermal quench is an important domain. For this purpose, {reliable and time efficient} multiphysics tools are required, which can provide valuable insights for the design of high-field magnets.

Electrothermal modeling has emerged as a powerful approach to analyze the complex thermal and electromagneic phenomena during a quench event in both hybrid and fully superconducting magnets with HTS REBCO inserts. By integrating the electrodynamic behavior of superconducting materials with the heat transfer processes, electrothermal models provide valuable insights into quench propagation, temperature distribution, and assessing role of the material properties from low temperatures {like 4.2 K} to room temperatures or above. These computational tools enable the optimization of magnet design, quench detection, and protection strategies, thereby enhancing the safety, performance, and reliability of advanced magnet systems. In recent years, various electrothermal modeling methods have been developed and employed, each offering unique advantages and considerations in capturing the intricacies of quench behavior in these advanced magnet systems. Most common numerical methods used in superconducting community for multiphysics quench modeling of are Finite Element Methods (FEM) \cite{LiJPCS2014, BadelSuST2019, dong2022APL, ChenSuST2023, stenvallA2023book, VitranoIEEE2023}, Finite Difference Methods (FDM) \cite{PISSANETZKYCry1994, BOTTURAJCP1996, JanitschkeIEEE2021, RavaioliIEEE2022}, and Lumped Electrical circuit model \cite{MaciejewskiMMAR2015, RAVAIOLICry2016, GarciaIEEE2017, bonnardCH2017SST, ChoIEEE2019, markiewicz2019SUST, gavrilin2021IEEE, genot2022IEEE, fazilleau2024IEEE}, for both electromagnetic and electrothermal analysis. {However,} these methods, {including FEM} commercial software like COMSOL and ANSYS, are quite slow in giving results for a whole magnet operation. {On the other extreme, simplified lumped models that consider non-insulated or metal-insulated pancake coils as a single unit can be very fast, but they lose detailed information on the in-coil processes \cite{bhattarai2020SUST}. Although several modelling works take radial currents in non-insulated and metal-insulated windings into account \cite{markiewicz2019SUST, ChoIEEE2019, gavrilin2021IEEE, dong2022APL, genot2022IEEE, fazilleau2024IEEE}, most of them ignore screening currents; which can cause an essential contribution in the dissipation in high-field magnets, specially during magnet charging or sudden discharge. Indeed, heat from screening currents may cause quench way before the operating current of a superconducting magnet is reached, as critical current decreases due to the rise in temperature. \R{Furthermore, the screening currents can affect the system mechanically as well, with high Lorentz pressure reducing turn-to-turn contact forces to zero at part of the winding \cite{yan2021SUST, li2022SUST, Srivastava2024SUST}. In this regions, the electrical and thermal conductance in the radial direction will be strongly suppressed, if not completely vanished; which might compromise the quench protection benefits of metal-insulated or non-insulated coils.} Remarkably, \cite{dong2022APL} considers screening currents, but only for a single pancake. In addition, \cite{gavrilin2024IEEE} partially takes screening currents into account for a full multi-pancake REBCO insert, but only regarding the change in critical current due to mechanical bending. Indeed, the AC loss considered in \cite{gavrilin2024IEEE, gavrilinAV2013IES} is by analytical fitting of magnetization measurements on tapes, and hence it seems to neglect magnetic shielding by screening currents, transport currents, and non-uniform dissipation within each tape.} Thus, a fast and robust multiphysics software {that can quickly analyze the quench effects with screening currents is required}.

{In this work, we present a novel, fast, and accurate electrothermal modeling method that takes both radial and screening currents into account in non-insulated (NI) and metal-insulated (MI) REBCO windings. This method is based on coupling the Minimum Electro-Magnetic Entropy Production (MEMEP) method \cite{pardoE2016SST, pardoE2024SSTa}, which is much faster than commercial codes, and our implementation of a finite difference method. We apply this method to a preliminary design of a 32 T magnet of EU project SuperEMFL \cite{superemfl2023}, which also aims to design a 40 T magnet. Here, we focus on electrothermal quench occurring during magnet ramp-up caused by either inefficient cooling or accidentally damaged turns. The aim of this numerical study is to evaluate the capability to avoid electrothermal quench entirely due to contact resistance between turns. For this reason, this article considers a wide range of contact resistance values. In the future, the developed software for quench analysis can provide researchers with an effective tool for a comprehensive study of quench phenomena in hybrid and HTS high field magnets. Thus it serves as a valuable resource to predict quench-related risks and mitigate them by optimizing the magnet design. The results of this paper were presented at Magnet Technology and EUCAS conferences in 2023, but not yet published \cite{dadhich2023MT}.
}

\begin{figure}[tbp]
	\centering

	\subfloat[][]
	{\includegraphics[trim=0 0 0 0,clip,width=10 cm]{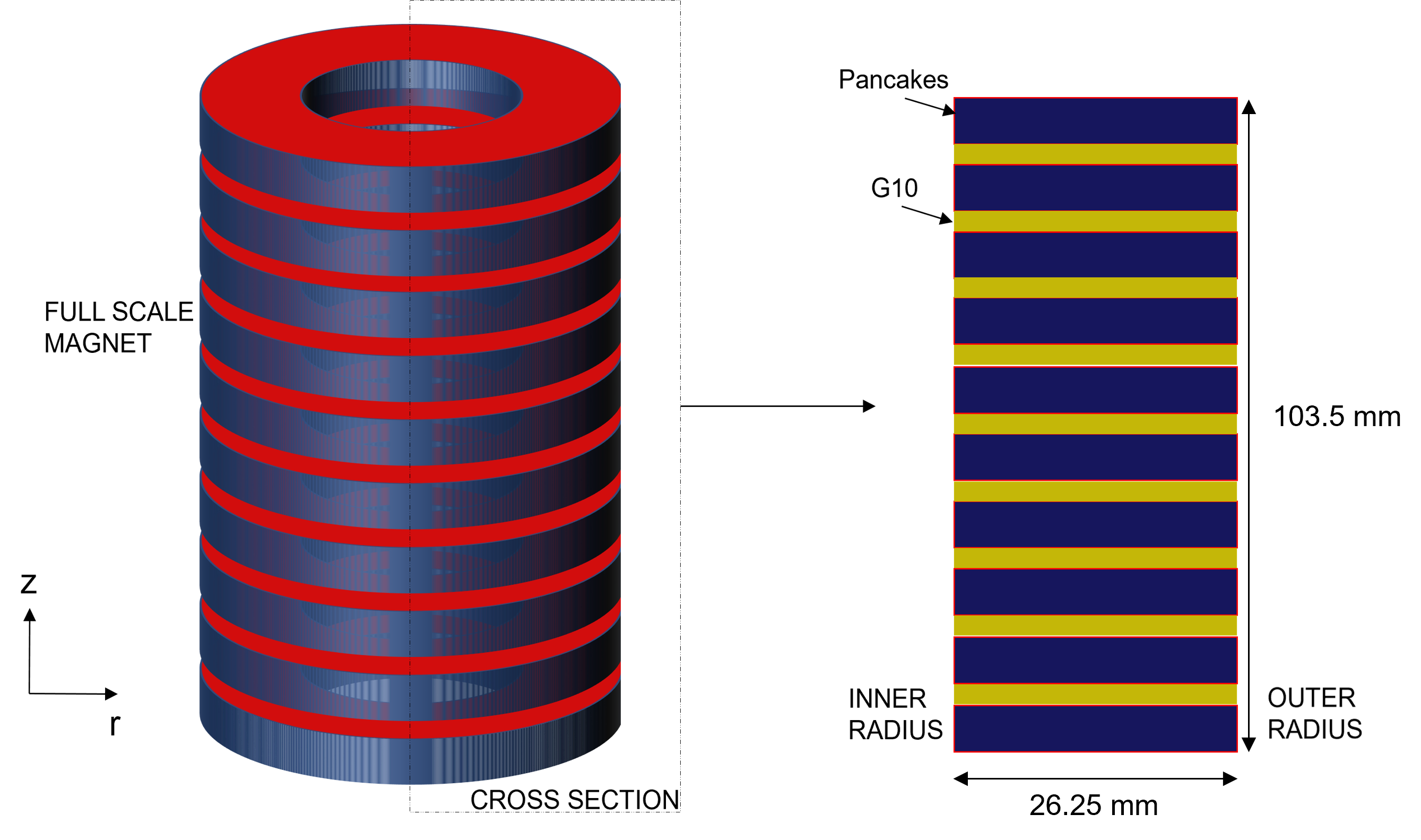}}

	\subfloat[][]
	{\includegraphics[trim=0 0 0 0,clip,width=10 cm]{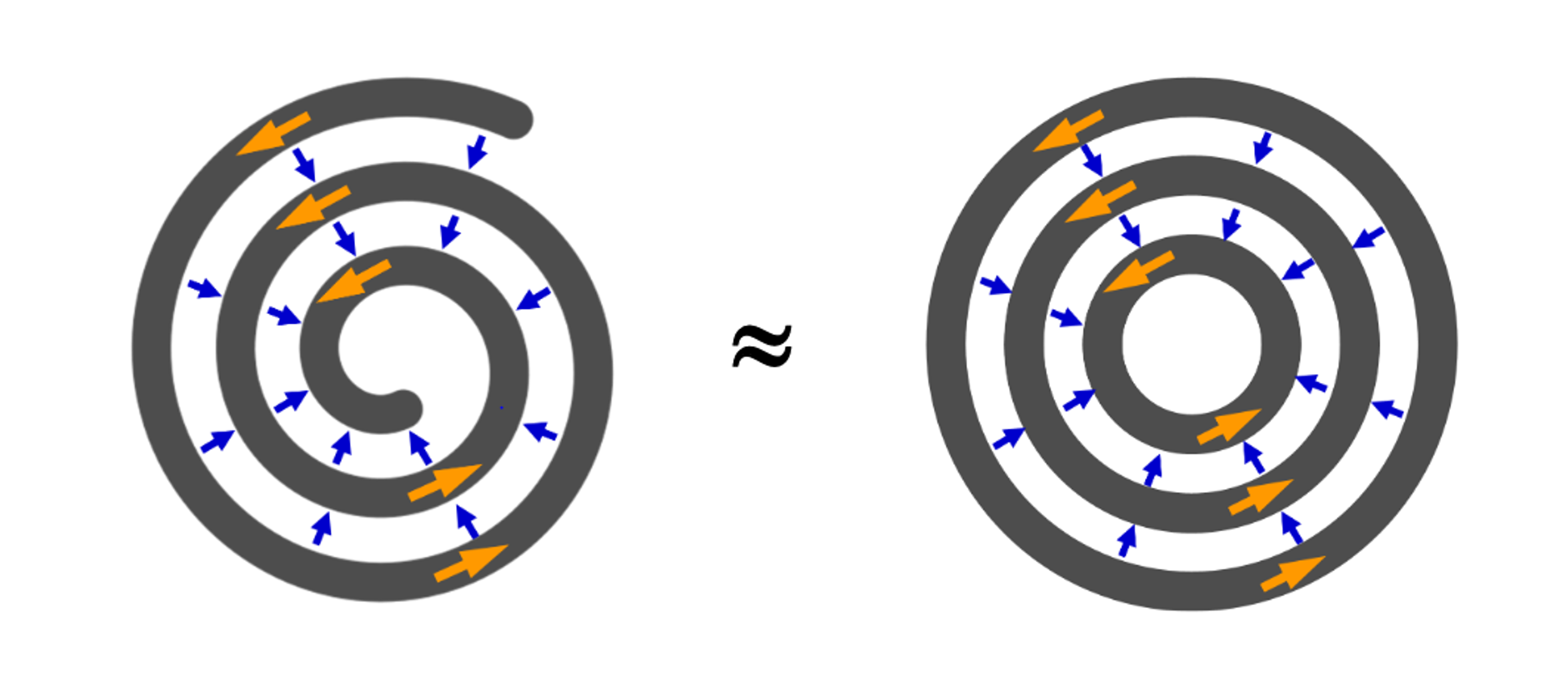}}

	\subfloat[][]
	{\includegraphics[trim=0 0 0 0,clip,width=10 cm]{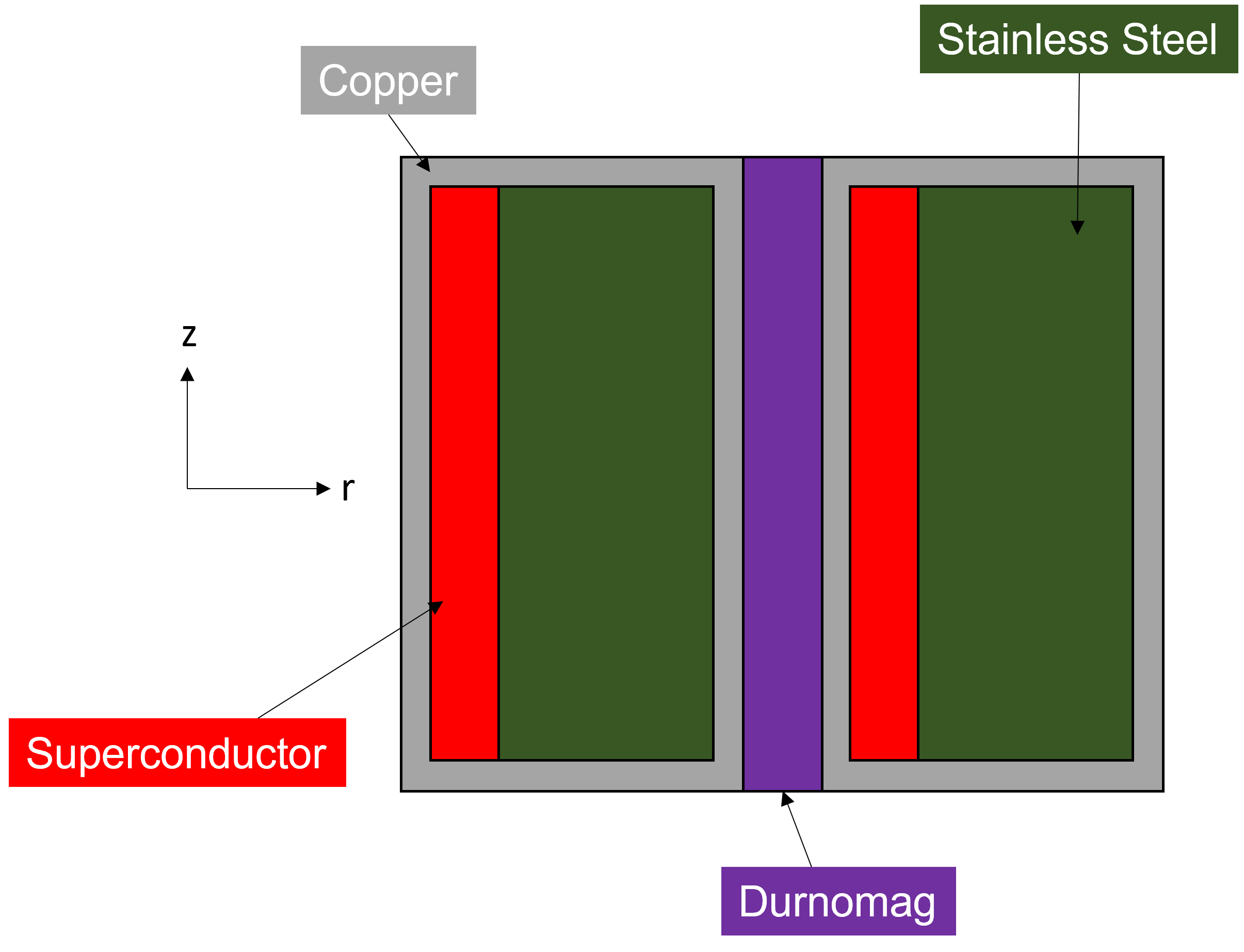}}

\caption{(a) Full scale magnet and its cross section. (b) Axisymmetric assumption for spiral pancakes (c) Cross section of turns in pancakes, including the isolating Durnomag$^{TM}$ layer between 2 tapes. The material layers in 'turns' or a single tape are homogenenized, which includes Durnomag$^{TM}$ as well.  }
\label{crossSec}
\end{figure} 

\section{Modeling Method}

The pancake coils in the magnet are spiral in nature, although we can take an axisymmetric assumption for the problem (Figure \ref{crossSec} (b)). Thus, the models considered in our software solve the Electromagnetic and Electrothermal equations in axisymmetric and cylindrical coordinates. These models are explained in detail below.

\subsection{Electromagnetic model}

We have used Minimum Electro-Magnetic Entropy Production (MEMEP) method, which is based on variational principles, for solving the electromagnetic part of the problem \cite{pardoE2015SST, pardoE2017JCP, pardoE2023book}. The benefit of using MEMEP for such problem is its usage of less degrees of freedom {compared to conventional finite-element methods}, which directly translates to faster calculations. {In addition, the parallel computing iterative algorithm in \cite{pardoE2016SST} further speeds up the computations for large windings. Recently, MEMEP has also been expanded in order to model non-insulated and metal-insulated REBCO inserts \cite{pardoE2024SSTa}. \R{An open source 3D version of the method's code is available on \cite{milancode}.}
}

{
As explained in \cite{pardoE2024SSTa}, we can approximate the spiral shape of each pancake by a set of concentric circular loops and a radial current flowing from loop to loop. The only assumptions of the simplified geometry are: the tape thickness plus isolation are much smaller than the coil inner radius; the eddy curents in the turn-to-turn currents are negligible and the turn-to-turn resistance is constant; and the magnetic field generated by the radial currents is negligible.

For simplicity, we assume that the electical properties of the REBCO tape are homogenized, which contains the superconducting layer, stabilization, and the metallic substrate \cite{pardoE2024SSTa}. This results in an equivalent orthotropic resistivity, which is non-linear. In order to speed-up the computations, we assume a further homogenized model, where each cross-sectional element contains several turns. This enables to model the electromagnetic properties of a REBCO insert in a highly time efficent way, being faster than real-time operation \cite{pardoE2024SSTa}. }

\subsection{Electrothermal model}

We have used explicit Finite Difference Method (FDM) {to solve} the thermal equations of the magnet. The benefit of using FDM is that it is very fast in its calculations, and very simple to code and implement. 

The {thermal diffusion} equation {that} is solved in FDM, is 
\begin{equation}
 { \rho_m \cdot C_p(T) \frac{\partial T}{\partial t} = \nabla \cdot (\ten{k}(T) \nabla T) + p(T)   },
\label{thermal}
\end{equation}
where, $\rho_m$, $C_p$, $T$, $t$, {$\ten{k}$}, and {$p$} are {the} mass density, {specific thermal capacity at constant pressure}, temperature, time, thermal conductivity {tensor}, and heat dissipation, respectively. {In our modelling, we assume that the thermal expansion is negligible, and hence the heat capacity at constant pressure, $C_p$, or at constant volume, $C_v$, are the same, $C_v \approx C_p$.} In addition, we use thermal capacity {per unit volume, $c_v$} = $\rho_m C_p$, in our model. Here, {$\ten{k}$} and  {$c_v$} are temperature dependent.  {Thus} Equation (\ref{thermal}) in cylindrical coordinates, assuming  {that $T$ does not depend on the anguar coordinate and that the thermal conductivity tensor is orthotropic}, is
\begin{equation}
 { c_v(T) \frac{\partial T}{\partial t} =   \frac{1}{r} \frac{\partial }{\partial r} (r \cdot k_r(T) \frac{\partial T}{\partial r}) + \frac{\partial }{\partial z} ( k_z(T) \frac{\partial T}{\partial z}) + p(T)   },
\label{thermaleq}
\end{equation}
where, $r$ and $z$ are the radial and axial coordinates of the magnet, respectively. Note that we consider different conductivities, $k_r$ and $k_z$, in radial and axial directions, respectively.

{Next, we discretize (\ref{thermaleq}) by assuming that $T$ is uniform in cells of rectangular cross-section and that the normal component of $\nabla T$ on the element surfaces is uniform. The cells are identified by two indexes in the $r$ and $z$ directions, respectively, $i,j$. We label the surfaces of each cell $i,j$ by shifting $\half$ of one index. The size of each cell in the $r$ and $z$ directions is $\Delta r_{i,j}$ and $\Delta z_{i,j}$, respectively. The resulting discretized thermal diffusion equation becomes
\begin{eqnarray} \label{Tdifdispar}
c_{v,i,j}(\partial_t T)_{i,j} & = & \frac{1}{\Delta r_{i,j}}\left [ \frac{r_{i+\half,j}}{r_{i,j}} k_{r,i+\half,j}(\partial_r T)_{i+\half,j} - \frac{r_{i-\half,j}}{r_{i,j}}k_{r,i-\half,j}(\partial_r T)_{i-\half,j} \right ] \\
&& + \frac{1}{\Delta z_{i,j}} \left [ k_{z,i,j+\half}(\partial_z T)_{i,j+\half} - k_{z,i,j-\half}(\partial_z T)_{i,j-\half} \right ] + p_{i,j} ,
\nonumber
\end{eqnarray}
where 
\begin{eqnarray}
(\partial_r T)_{i+\half,j}=\frac{T_{i+1,j}-T_{i,j}}{r_{i+1,j}-r_{i,j}} \\
(\partial_r T)_{i-\half,j}=\frac{T_{i,j}-T_{i-1,j}}{r_{i,j}-r_{i-1,j}}
\end{eqnarray}
and $(\partial_z T)_{i,j+\half}$, $(\partial_z T)_{i,j-\half}$ are calculated equivalently. In (\ref{Tdifdispar}),} terms similar to $k_{i+\frac{1}{2}, j}$ are effective thermal conductivities at the surface between two cells. Considering the $k_{i+\frac{1}{2}, j}$ case, {we calculate this effective conductivity at the cells surface using that the heat flux from the center of one cell to its neighbour in the $r$ direction experiences a heat resistance
\begin{equation} \label{Rrsur}
R_{i+\frac{1}{2},j}=\frac{1}{2}R_{i,j}+\frac{1}{2}R_{i+1,j}
\end{equation}
and that the effective conductivity is related to the heat resistance as
\begin{equation} \label{Rrsurk}
R_{i+\frac{1}{2},j}=\frac{ \frac{1}{2}(\Delta r_{i,j} + \Delta r_{i+1,j}) }{\Delta z_{i,j}\cdot 2\pi (r_{ij}+\frac{1}{2}\Delta r_{i,j} )k_{i+\frac{1}{2},j}} ,
\end{equation}
where the top of the denominator is the length that the heat flux travels and the term before $k_{i+\half,j}$ at the denominator is the average cross-section that the heat flux experiences. Assuming that $\Delta r_{ij}$, $\Delta r_{i+1,j} \ll r_{ij}$ and using (\ref{Rrsur}) and (\ref{Rrsurk}), we obtain
\begin{equation} \label{kbody}
k_{i+\frac{1}{2},j}=(\Delta r_{ij} + \Delta r_{i+1,j}) \left [ \frac{ \Delta r_{ij} } { k_{r,ij} } + \frac{ \Delta r_{i+1,j} } { k_{r,i+1,j} } \right]^{-1} .
\end{equation}
The effective surface conductivity in the $z$ direction, for the surfaces at $(i,j+1/2)$, is found in the same way, obtaining
\begin{equation}
k_{i,j+\frac{1}{2}}=(\Delta z_{ij} + \Delta z_{i,j+1}) \left [ \frac{ \Delta z_{ij} } { k_{z,ij} } + \frac{ \Delta z_{i,j+1} } { k_{z,i,j+1} } \right ]^{-1}.
\end{equation}.
}

{
In our model, we also take cooling by a cryogenic fluid like liquid helium into account by introducing an effective heat conductivity that causes the same heat flux, as follows. First, the heat flux flowing outwards from the object is
}
\begin{equation}
G_s = h(T_s) \cdot (T_s - T_l), 
\label{Gs}
\end{equation} 
where, $T_l$ is the temperature of liquid coolant (4.2 K in our case), and $T_s$ is the temperature of the surface, and $h$ is the heat convection coefficient. {Additionally, for the surface at the outer radius of the whole sample, the heat flux towards outside the coil is
\begin{equation} \label{Gijsur}
G_{i+\half,j}=-k_{r,i,j}\frac{T_s-T_l}{\Delta r_{ij}/2} .
\end{equation}
Equating $G_{i+\half,j}$ with $G_s$ above leads to
\begin{equation} \label{Tsur}
T_s= \frac{ h(T_s)+\frac{2k_{r,i,j}}{\Delta r_{i,j}}T_{i,j} }{ h(T_s)+\frac{2k_{r,i,j}}{\Delta r_{i,j}} }
\end{equation}
To exactly obtain $T_s$, we should solve this equation numerically for the given non-linear $h(T_s)$. In this work, we assume $h(T_s)\approx h(T_{i,j})$ for simplicity, which will be achieved when $\Delta r_{i,j}$ is sufficiently small. Next, we obtain the effective thermal conductivity at the surface, $k_{i+\half,j}$, that we define such that
\begin{equation}
G_{i+\half,j}=-k_{i+\half,j}\frac{T_l-T_{i,j}}{r_{i+1,j}-r_{i,j}},
\end{equation}
where we added a layer of elements on the liquid, just beyond the outer radius with constant temperature $T_l$. Equating to (\ref{Gijsur}) and using (\ref{Tsur}), we obtain
\begin{equation} \label{ke}
k_{i+\half,j}=\frac{h(T_s)\cdot(r_{i+1,j}-r_{ij})}{1+\frac{2h(T_s)\Delta r_{i,j}}{k_{r,i,j}}}.
\end{equation}
In the same way, we obtain the surface resistivities at the inner radius and the top and bottom surfaces of the winding. Then, for the element surfaces of constant $r$, we apply either (\ref{ke}) or (\ref{kbody}) case wise, depending whether the elemement surface belongs to the object boundary or the body, respectively. We evaluate $k_{i-\half,j}$, $k_{i,j+\half}$, $k_{i,j-\half}$ equivalently.

\R{We do not consider any additional interface thermal resistance between layers, since these values are not experimentally available. The model can handle an additional interface thermal resistance in the radial direction, if required.}

}

{
Finally, we find the temperature distribution at time $n+1$ from that at time $n$ by direct Euler integration of (\ref{Tdifdispar}), and hence $(\partial_t T)_{ij} \approx (T_{ij}^{n+1}-T_{ij}^n)/\Delta t$. Expanding the partial derivatives in space, we obtain
\begin{eqnarray} \label{FDM}
 T_{ij}^{n+1} & = & T_{ij}^n + \frac{\Delta t}{c_{v,i,j}^n\Delta r_{i,j}} \left [ \frac{r_{i+\half,j}}{r_{i,j}} k_{r,i+\half,j} \frac{T_{i+1,j}^n-T_{i,j}^n}{r_{i+1,j}-r_{ij}} - \frac{r_{i-\half,j}}{r_{i,j}}k_{r,i-\half,j}\frac{T_{i,j}^n-T_{i-1,j}^n}{r_{i,j}-r_{i-1,j}}
\right ] \nonumber \\
 && + \frac{\Delta t}{c_{v,i,j}^n\Delta z_{i,j}} \left [ k_{z,i,j+\half} \frac{T_{z,i,j+1}^n-T_{z,i,j}^n}{z_{i,j+1}-z_{i,j}} - k_{z,i,j-\half} \frac{T_{z,i,j}^n-T_{z,i,j-1}^n}{z_{i,j}-z_{i,j-1}} \right ] \nonumber \\
 && + \frac{\Delta t p_{i,j}^n}{c_{v,i,j}} .
\end{eqnarray}
}
{Thus, equation (\ref{FDM})} depicts the general FDM equation we use in our Electrothermal solver. Notably, the equation is able to consider temperature dependent $c_v$ and $\ten{k}$, and also variable mesh size, which is important when we introduce other external materials to the system, such as G10 between pancakes. This allows us the freedom to choose different mesh element sizes for superconducting turns and other materials.  

The stability condition for the FDM method is
\begin{equation}
 \Delta t \leq \frac{1}{2} \frac{\Delta r^2 \Delta z^2}{\Delta r^2 + \Delta z^2} \cdot \frac{{\rm min} (C_v)}{{\rm max} (k)},  
\label{stability}
\end{equation}
which shows that the required number of timesteps for a stable solution depends on mesh element size, and thermal properties. With non-linear situations, like HTS magnets, the thermal conductivity and heat capacity can vary along the mesh, and thus it is required to choose their maximum and minimum values to satisfy this condition. 

\subsection{Coupling of methods}

\begin{figure}
	\centering
	{\includegraphics[trim=0 0 0 0,clip,width=13 cm]{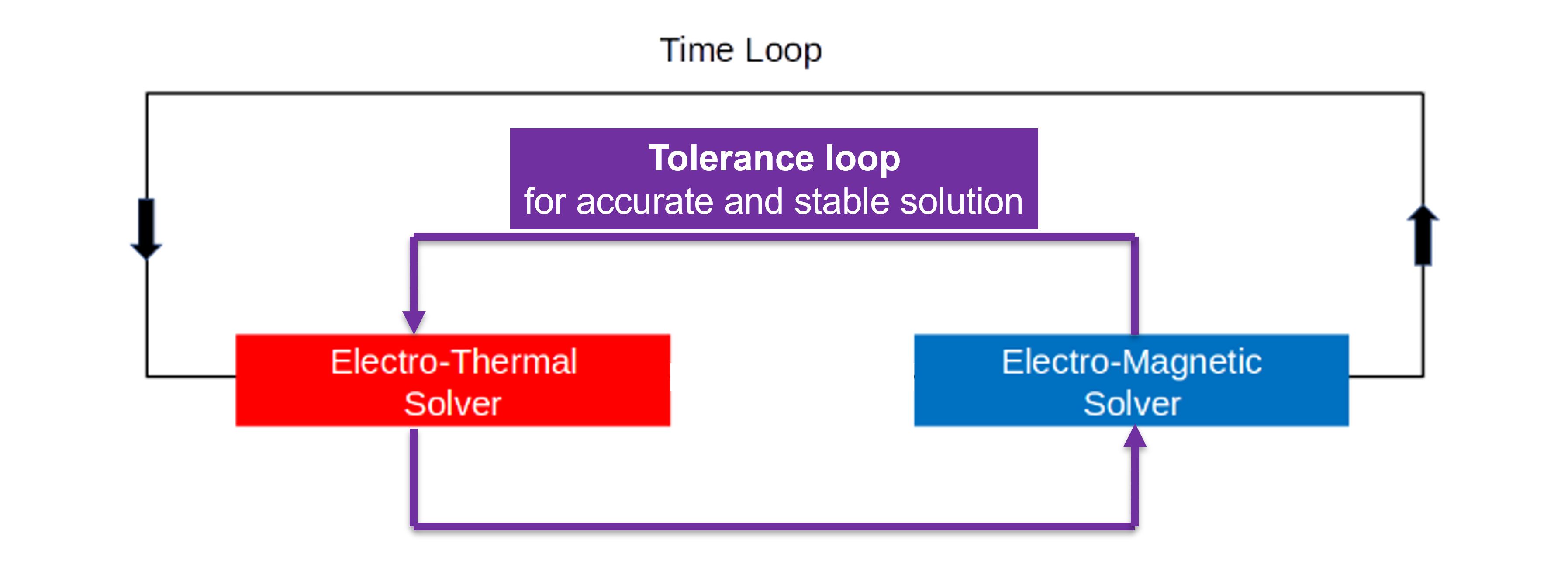}}
\caption{ Coupled Electromagnetic (MEMEP) and Electrothermal (FDM) solvers. Internal iterative tolerance loops are added to improve accuracy of solutions.  } 
\label{solver}
\end{figure}

The Electromagnetic (MEMEP) and Electrothermal (FDM) solvers are coupled, and provide input for each other in an iterative manner, to give a complete coupled software. The in-house Electromagnetic and Electrothermal solvers are written in Fortran and C++ languages respectively, and they communicate with each other directly using a combination of Fortran and C++ wrappers. Other way for them to communicate is through reading the solution files, which is not considered here as it can significantly slow down the calculations, and wrappers are used for fast and direct internal communication of these models. The input for FDM is the power generated from MEMEP solutions, and the temperature solutions serve as an input for MEMEP and $J$ solutions, as we consider complete $J_c(B,T)$ dependence and temperature dependence of the metal resistivities. Also, the models are run through a tolerance loop for accurate solutions at each time step. This coupling is shown in Figure \ref{solver}.

To fasten our calculations, we calculate MEMEP on longer timesteps than FDM. Between two MEMEP timesteps, we use linear interpolation of the power solutions, to provide input for FDM on the remaining timesteps. As seen in Results section, the power curves are continuous, and quite linear in most cases, so this approximation is valid. Thus, this combination of MEMEP and FDM gives rise to an effective and fast tool for quench simulations.

\subsection{Magnet configuration and material properties}

\begin{figure*} [tbp]

\begin{multicols}{2}    
		\centering
	\subfloat[][]
	{\includegraphics[trim= 0 0 0 0, clip, width = 8 cm]{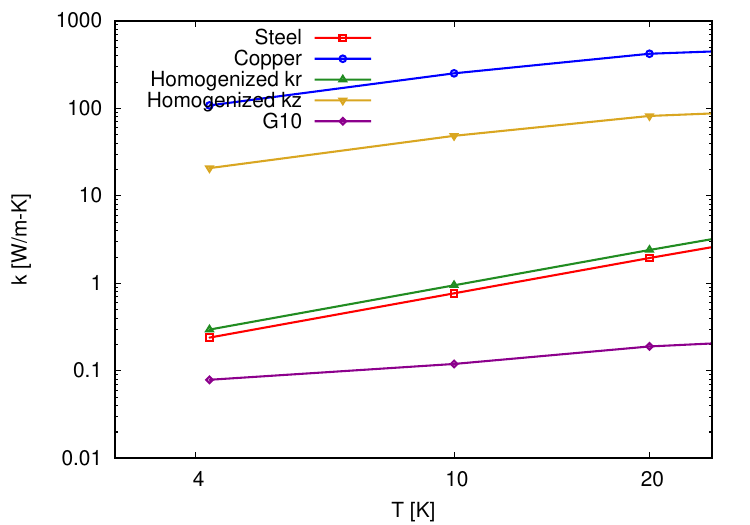}}
	\subfloat[][]
	{\includegraphics[trim= 0 0 0 0, clip, width = 8 cm]{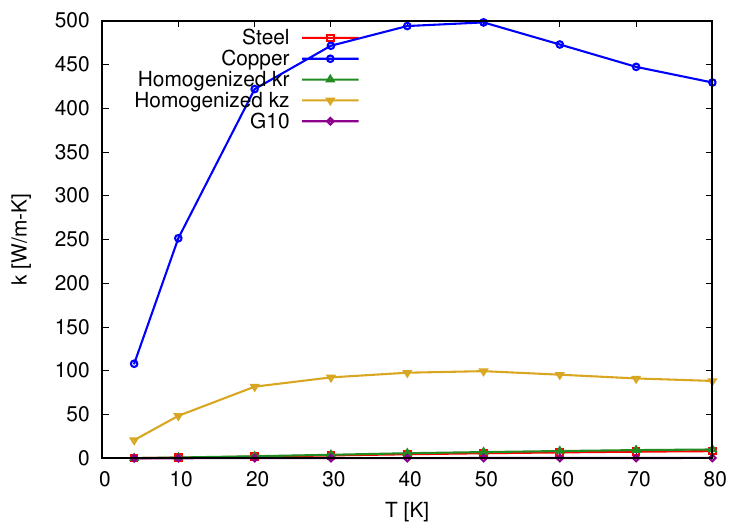}}	 
\end{multicols}

\begin{multicols}{2}    
		\centering
	\subfloat[][]
	{\includegraphics[trim= 0 0 0 0, clip, width = 8 cm]{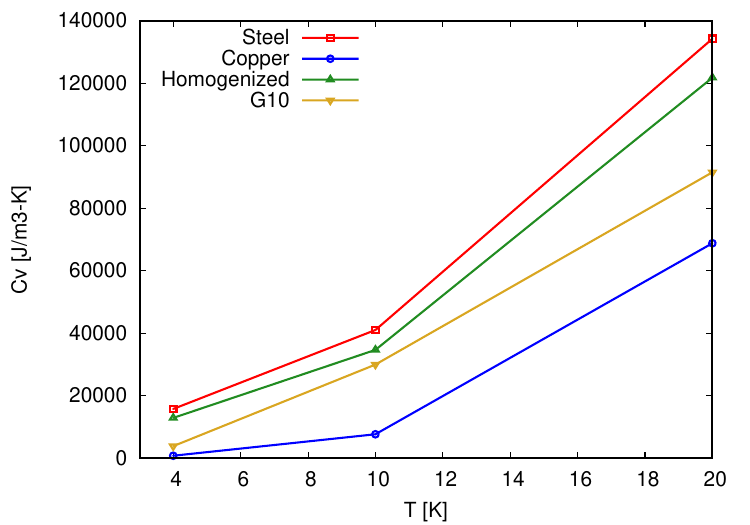}}
	\subfloat[][]
	{\includegraphics[trim= 0 0 0 0, clip, width = 8 cm]{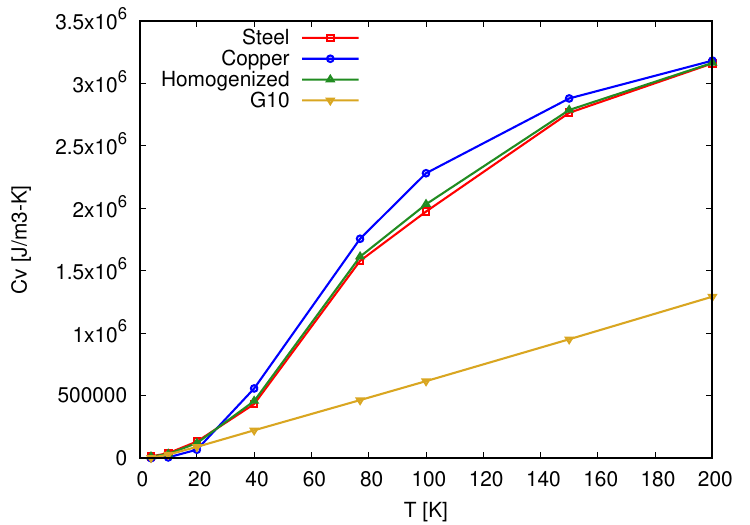}}	 
\end{multicols}

		\centering
	\subfloat[][]
	{\includegraphics[trim= 0 0 0 0, clip, width = 10 cm]{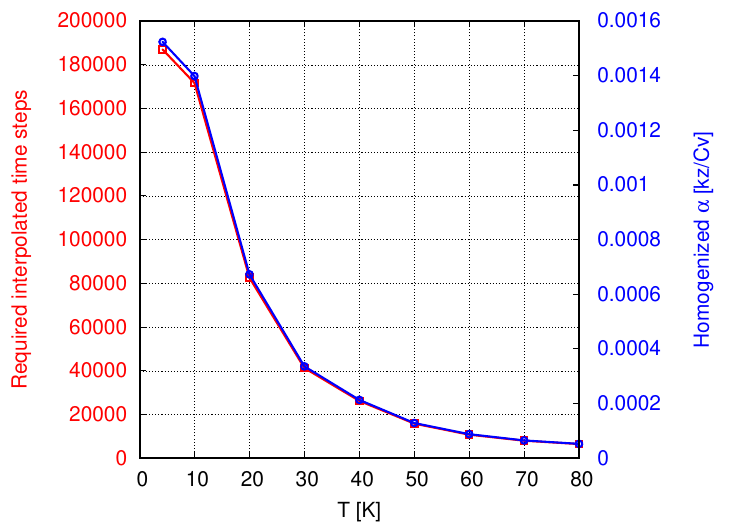}}

\caption{ (a) and (b) shows thermal conductivities of layers at low temperatures and high temperatures, respectively. Similarly, (c) and (d) shows thermal capacities at low and high temperatures. (e) shows that the required timesteps for FDM reduces as temperature increases, which is due to decrease in $\alpha$.  } 
\label{props}
\end{figure*}

We have considered 16 pancakes in a magnet for our calculations. G10 material is present between the pancakes for heat exchange within pancakes, as seen in the magnet cross section in Figure \ref{crossSec} (a). The pancakes in the magnet are made of REBCO tapes or turns, and the tape configuration considered is shown in Figure \ref{crossSec} (c). We consider complete homogenization of the electrical and thermal properties of layers in superconducting tapes for our methods, as we have shown in our previous work \cite{PardoIEEE2023}. As a consequence, the heat conductivity in the $z$ direction is much higher than in the $r$ direction. The reason is that we consider a copper layer around the tape for thermal stability, which increases the thermal conductivity in the z direction. In addition, homogenization of a tape includes the isolating layer, or Durnomag$^{TM}$ material in our considered metal-insulated coil, which reduce the thermal conductivity in the $r$ direction. We ignore the silver layer in our case, as it does not influence the thermal solutions much due to its comparatively small thickness (few microns). We also consider stainless steel properties for Hastelloy, HTS layer, and Durnomag$^{TM}$, as they are similar alloys and the data is readily available, and the small differences in values does not significantly affect quench. In addition, the assumed thermal properties for the superconductor will cause little impact, due to its small thickness, We also take different thermal conductivities in both radial and axial directions \cite{PardoIEEE2023}.  These material properties are shown in Figure \ref{props}.

Figure \ref{props} (a)-(d) shows stark difference in the thermal properties of different layers. We see that the thermal conductivity and thermal capacity of copper is orders of magnitude higher and lower, respectively, when compared to the other layers, especially at low temperatures (below 20 K). Due to this low thermal capacity, the HTS tapes heat up very fast at this low temperature range, which leads to fast thermal quench as we see in Results section later. The high thermal conductivity of copper also leads to fast propagation of quench across the tape width or in the angular direction.

We also see that the temperature dependent thermal diffusivity ($\alpha = k_z/c_v$) in Figure \ref{props} (e) is very high at low temperatures. Here we consider conductivity in axial ($z$) direction for plotting and determination of time steps, as it is higher than radial conductivity, and the maximum thermal conductivity in the system matters for FDM (Equation \ref{stability}). Due to that, the number of required interpolated timesteps for FDM (between 2 timesteps of MEMEP) is much higher at lower temperatures. But as the temperature of the system rises, $\alpha$ decreases, and thus the required number of timesteps is much lower at higher temperatures (above 20 K). Thus, we show that explicit FDM is highly advantageous for applications in high temperature environment, due to less required timesteps leading to faster calculations. Modern computers can still use FDM for low temperatures, as they are getting faster with improvement in technology. Our software is also able to consider variable number of timesteps required for stable solutions, and calculates it automatically as the temperature rises. {Because of that, most calculations in this text take less than 12 hours using a normal desktop, mainly due to fast electromagnetic calculations. These computation times may be improved in future using parallel computing for FDM and other optimization techniques. Moreover, MEMEP is benchmarked various times before, as well as, the coupling between MEMEP and FDM (for racetrack coils) \cite{pardoE2023book, PardoIEEE2023}. We plan to compare the results of this research in future with experiments and other methods in our project like PEEC model, and publish, as the project evolves \cite{fazilleau2024IEEE, superemfl2023}.}

\begin{figure}
	\centering
	
	{\includegraphics[trim=0 0 0 0,clip,width=10 cm]{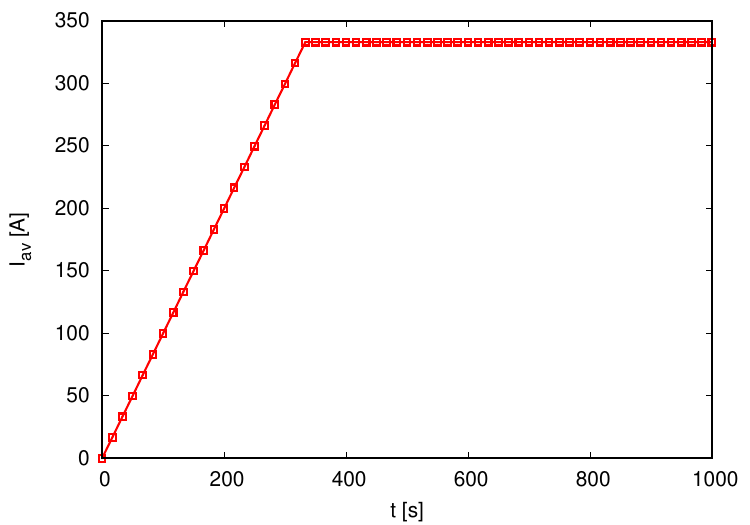}}
\caption{ For the input current, we use a ramp of 1 A/s followed by a long plateau. } 
\label{current}
\end{figure}

\begin{figure}
	\centering
	
	{\includegraphics[trim=0 0 0 0,clip,width=18 cm]{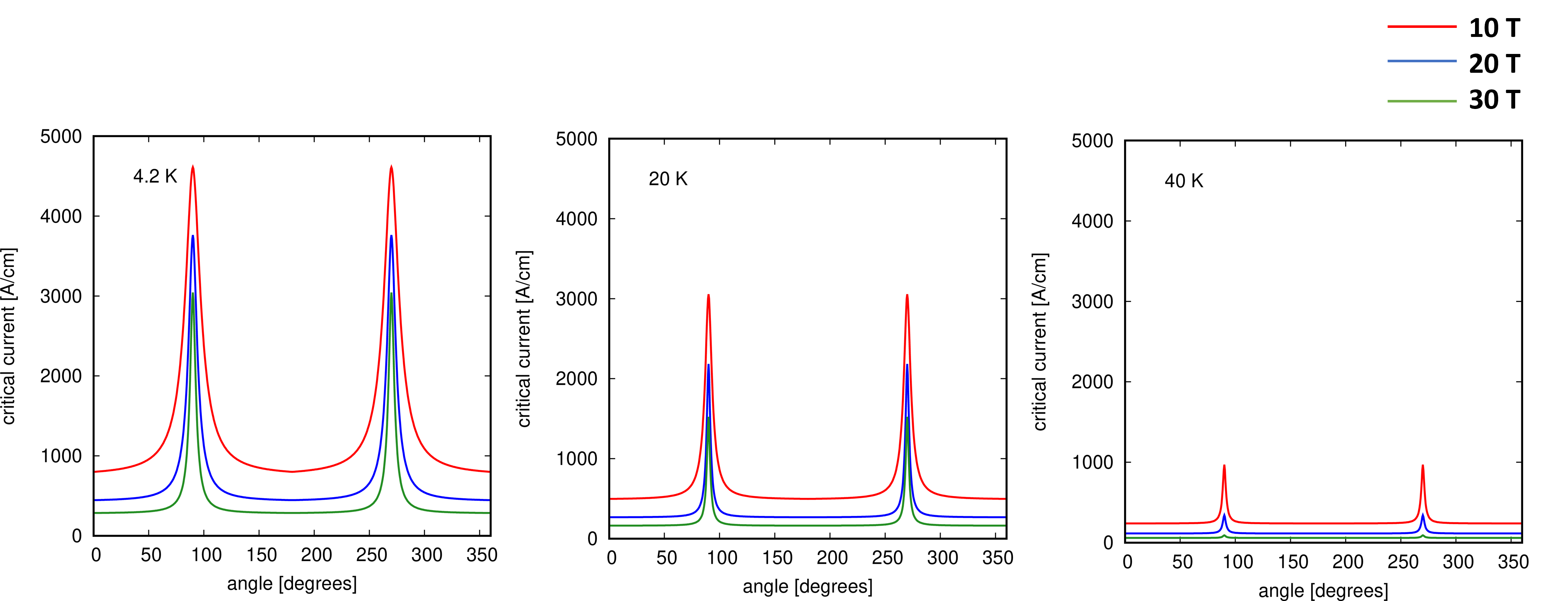}}
\caption{ J$_c$(B,T,$\theta$) dependence of Fujikura tape at different temperatures. } 
\label{JcB}
\end{figure}

As an input, we use a current ramp of 1 A/s, upto 333 seconds, and then a stable current plateau till a total of 1000 seconds, as seen in Figure \ref{current}. We use 20 timesteps for MEMEP during the current rise (up to 333 A), which leads to $\Delta t$ for MEMEP around 16 seconds for each time step. This time step is very large, which will require many timesteps for FDM. In addition, we consider a relatively fast ramp for a high-field magnet, and hence the instantaneous dissipation can be considerable.

\begin{table}
\centering
\caption{ {Magnet Configuration}}
\label{param1}
\footnotesize
\begin{tabular}{@{}llll}
\br

Material & Thickness [$\mu$m] &  Magnet dimensions  &   Values  \\
\mr
HTS         & 1         & Inner Diameter      &  25 mm    \\
Substrate   & 54        & Outer Diameter      &  51.25 mm   \\
Copper      & 10        & No. of pancakes     &  16   \\
Durnomag$^{TM}$    & 30        & No. of turns (per pancake)    & 250    \\
Conductor width & 6 mm    & G10 thickness           &  0.5 mm   \\
            &           & Background field    & 19 T     \\        

\br
\end{tabular}\\

\end{table}
\normalsize

\begin{figure}
			\centering
	\subfloat[][]
	{\includegraphics[trim=0 0 0 0,clip,width=10 cm]{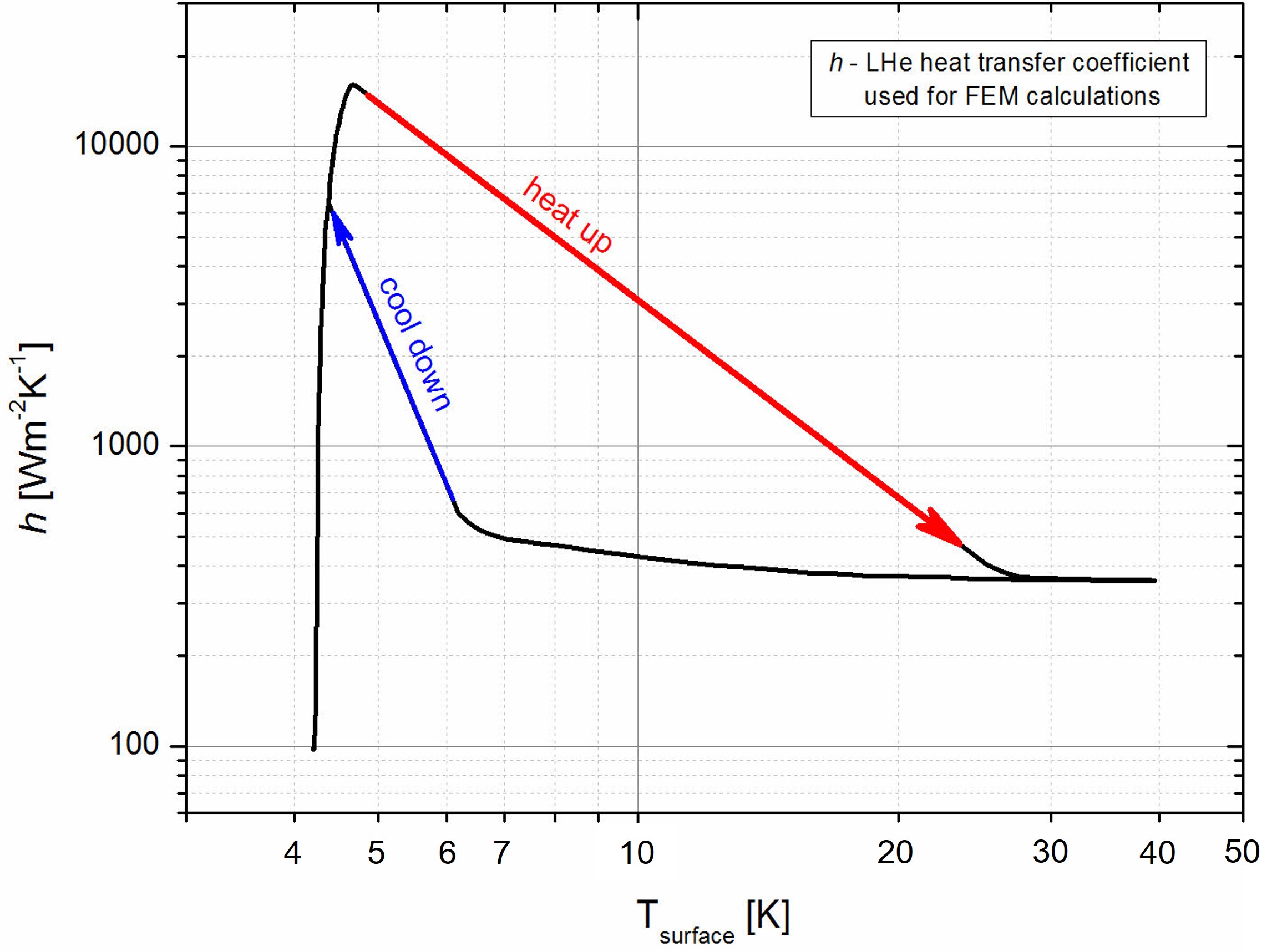}}
	
				\centering
	\subfloat[][]
	{\includegraphics[trim=0 0 0 0,clip,width=15 cm]{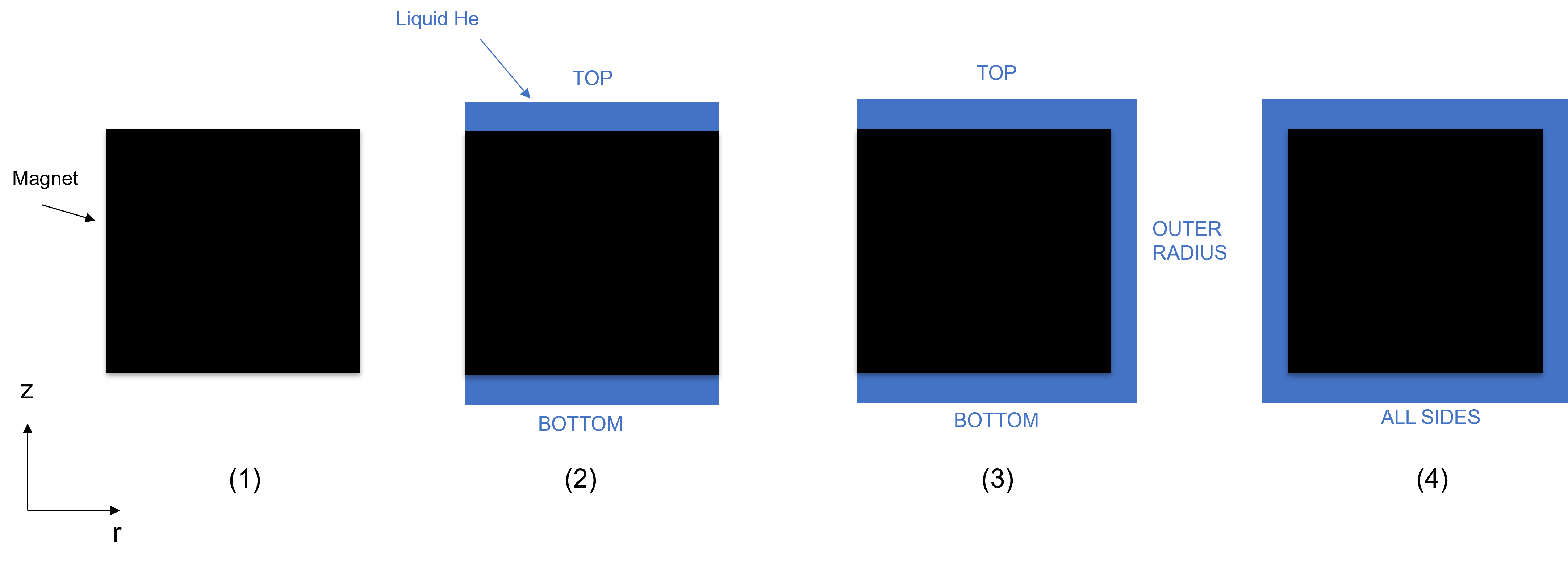}}
	
\caption{ (a) Convection coefficient curve for liquid Helium. We use 'Up' part of this curve currently. (b) shows different cooling configurations/boundary conditions of the magnet. (b)(1) is the adiabatic case with no cooling by external Helium.   } 
\label{hcurve}
\end{figure}

{The dimensions for the magnet and layers is given is Table \ref{param1}.} We consider perfect thermal conductor conditions at the interface between pancake and G10, mainly due to the lack of data at the current state-of-the-art, and hence we can directly use the temperature dependent equation \ref{kbody} at this surface. A full $J_c(B,T,\theta)$ dependence is used for the calculations, as shown in Figure \ref{JcB}. This $J_c(B,T,\theta)$ dependence corresponds to an analytical fit of measured data of Fujikura tape \cite{Fleiter2014}. \R{More details about the models and magnet data can be found at \cite{Pardo2024SUST}.} We also incorporate cooling of the magnet using liquid Helium, for which the convection coefficient ($h$) curve is shown in Figure \ref{hcurve} (a), of which we use 'heat up' part in this work \cite{van2015helium}. Different thermal boundary conditions for the magnet are shown in Figure \ref{hcurve} (b), which contributes to the cooling of magnet from different sides. Case 1 in this figure has no cooling from any side, which comes under presumed adiabatic conditions (i.e., assuming magnet is in vaccuum). This case is important to simulate, as it shows the undisturbed heating of the magnet, and hence without cooling. Additionally, the current density and temperature profiles don't show the current and temperature on G10 for simplicity, and only the pancakes are shown in the Results section.

\section{Results and Discussion}

\subsection {Adiabatic case} 

\begin{figure}
	
				\centering
	\subfloat[][]
	{\includegraphics[trim=0 0 0 0,clip,width=8 cm]{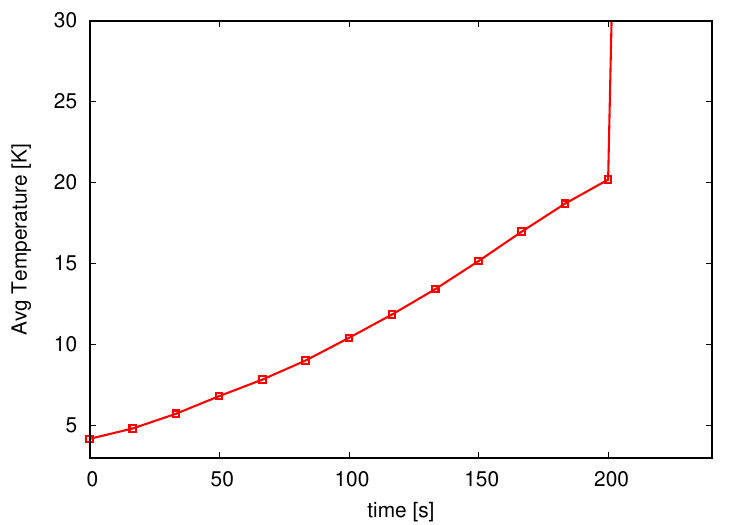}}
	
				\centering
	\subfloat[][]
	{\includegraphics[trim=0 0 0 0,clip,width=12 cm]{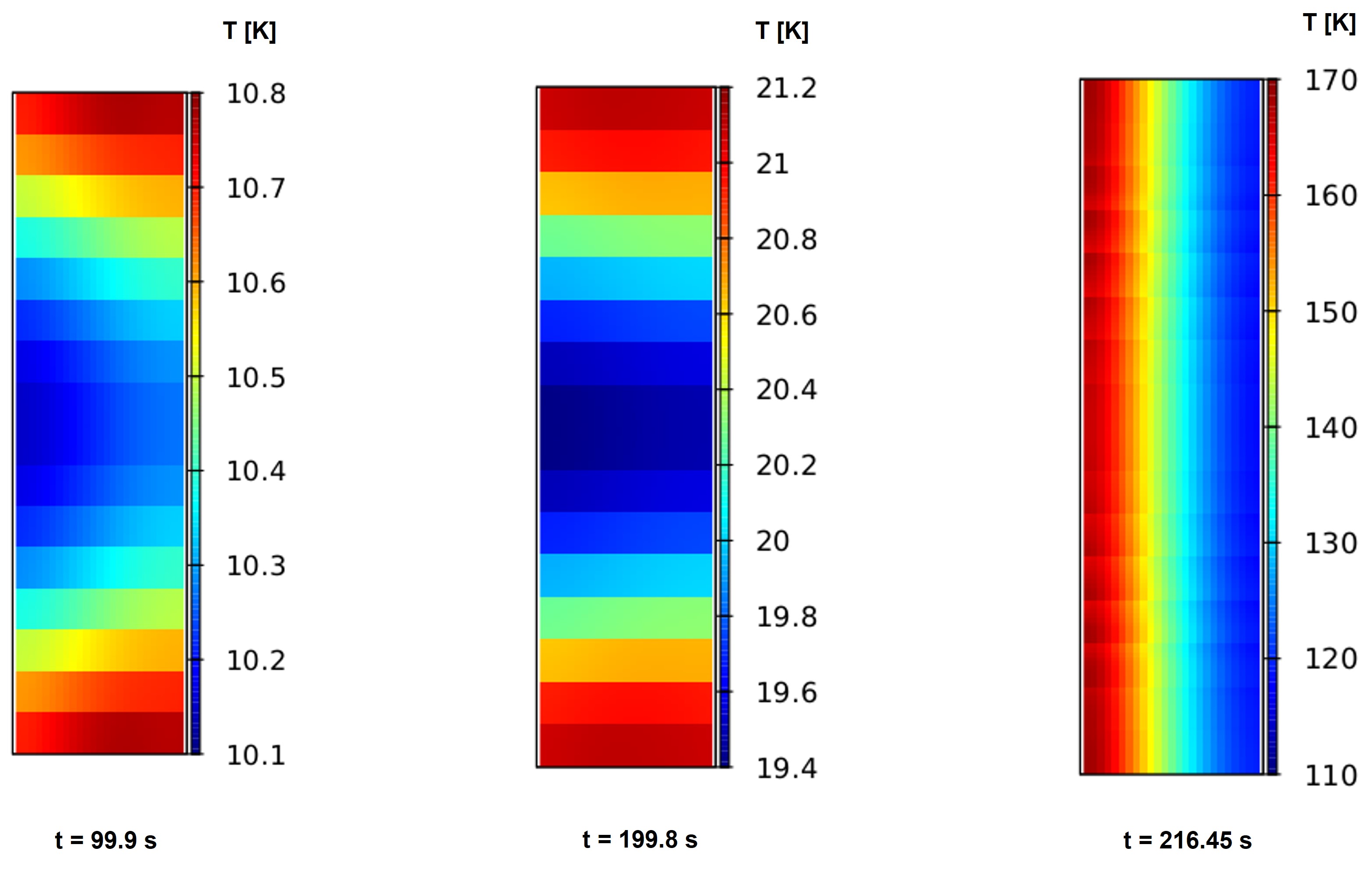}}
	
					\centering
	\subfloat[][]
	{\includegraphics[trim=0 0 0 0,clip,width=12 cm]{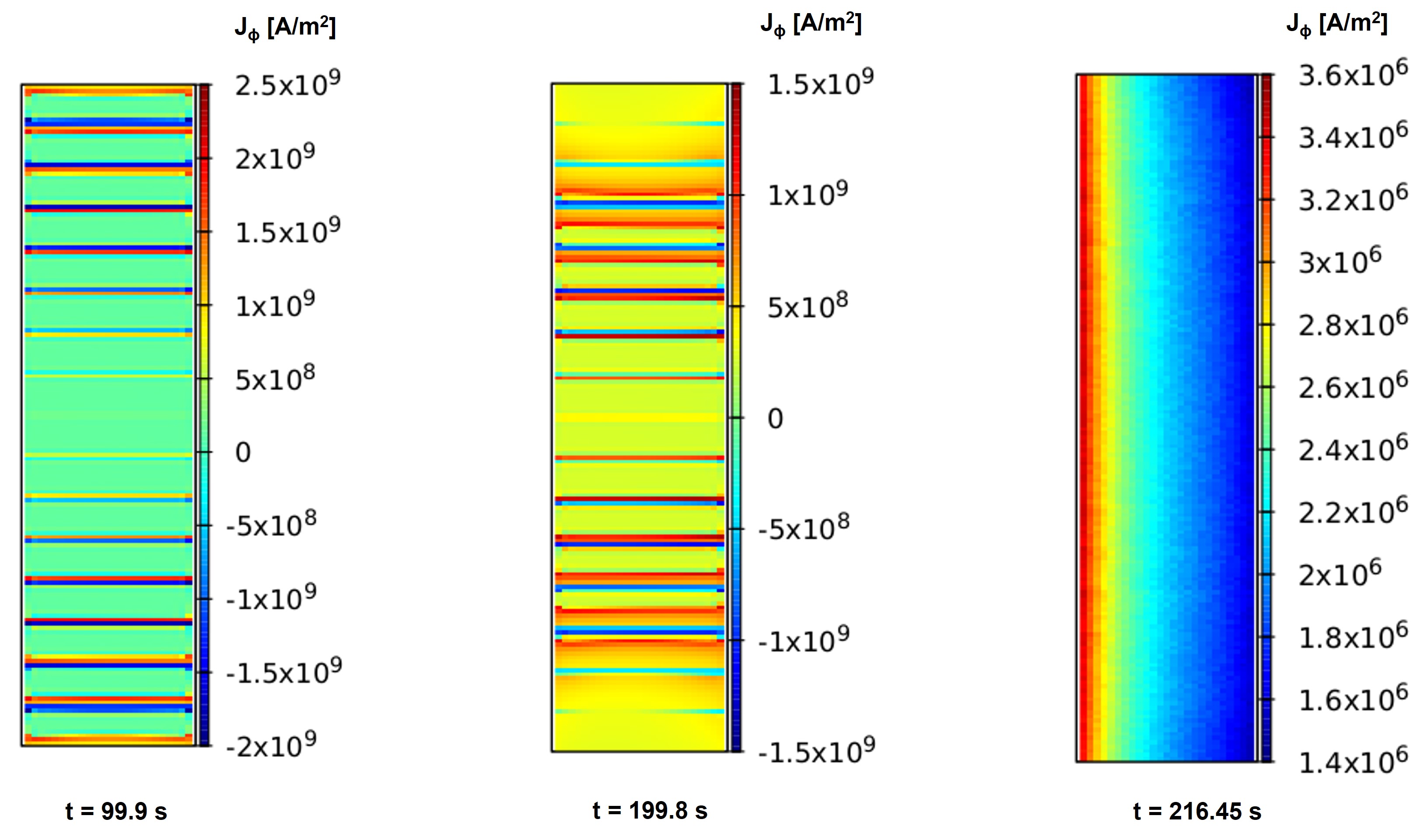}}
	
\caption{ Adiabatic case (a) Average temperature, (b) Temperature profiles, and (c) J profiles at different times. Adiabatic assumption is not enough to check the complete operation of the magnet as the magnet quenches prematurely.} 
\label{adia}
\end{figure}

The results for adiabatic condition (Figure \ref{hcurve} (b)(1)) is shown in Figure \ref{adia}. It can be seen in Figure \ref{adia} (a) that the average temperature of the magnet rises rapidly under these conditions and the magnet quenches between 199.8 s and 216 s. Thus, the magnet cannot be operated over the whole ramp (till 333 s) in adiabatic conditions, and hence proper cooling conditions are required to simulate the magnet run. Figure \ref{adia} (b) and (c) shows the temperature and current density profiles for this case, respectively. It can be seen here that the current starts penetrating from the top and bottom pancakes, which causes that the temperature rise is the highest in these sections. At 216 s, the magnet is quenched, and hence the profiles show 'normal' or non-superconducting behavior. This sharp rise in temperature is due to low thermal capacities of material layers in HTS tape at low temperatures (under 20 K). {It is also seen that initially the temperature is higher at outer radius (99 s), but later the temperature rise is higher at inner radius (216 s). This is due to the sudden rise in radial currents later at inner radius, which gives rise to higher power loss than at the outer radius.}

\subsection{Cooling cases} 

\begin{figure*} [tbp]

\begin{multicols}{2}    
		\centering
	\subfloat[][]
	{\includegraphics[trim= 0 0 0 0, clip, width = 8 cm]{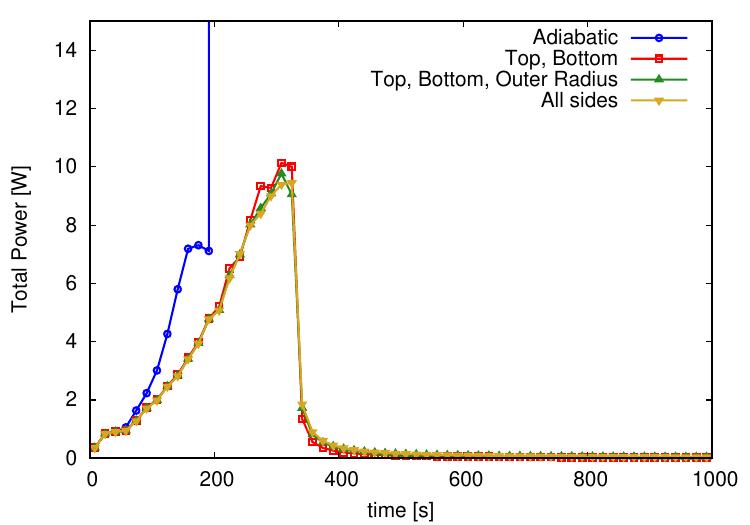}}
	\subfloat[][]
	{\includegraphics[trim= 0 0 0 0, clip, width = 8 cm]{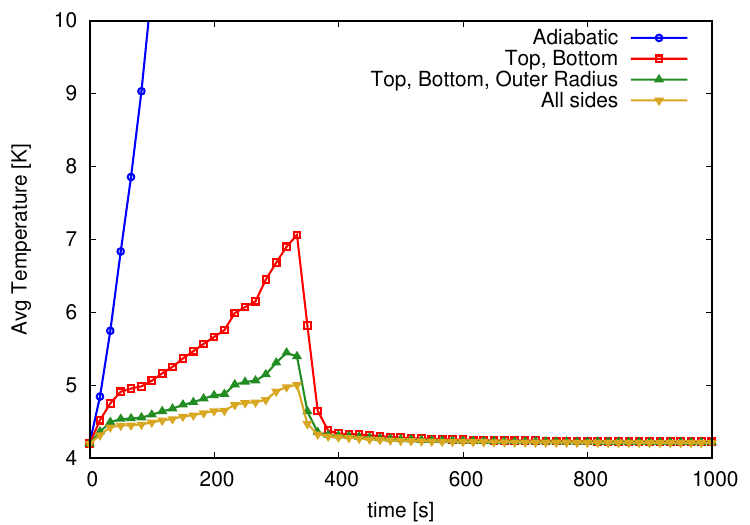}}	 
\end{multicols}

		\centering
	\subfloat[][]
	{\includegraphics[trim= 0 0 0 0, clip, width = 12 cm]{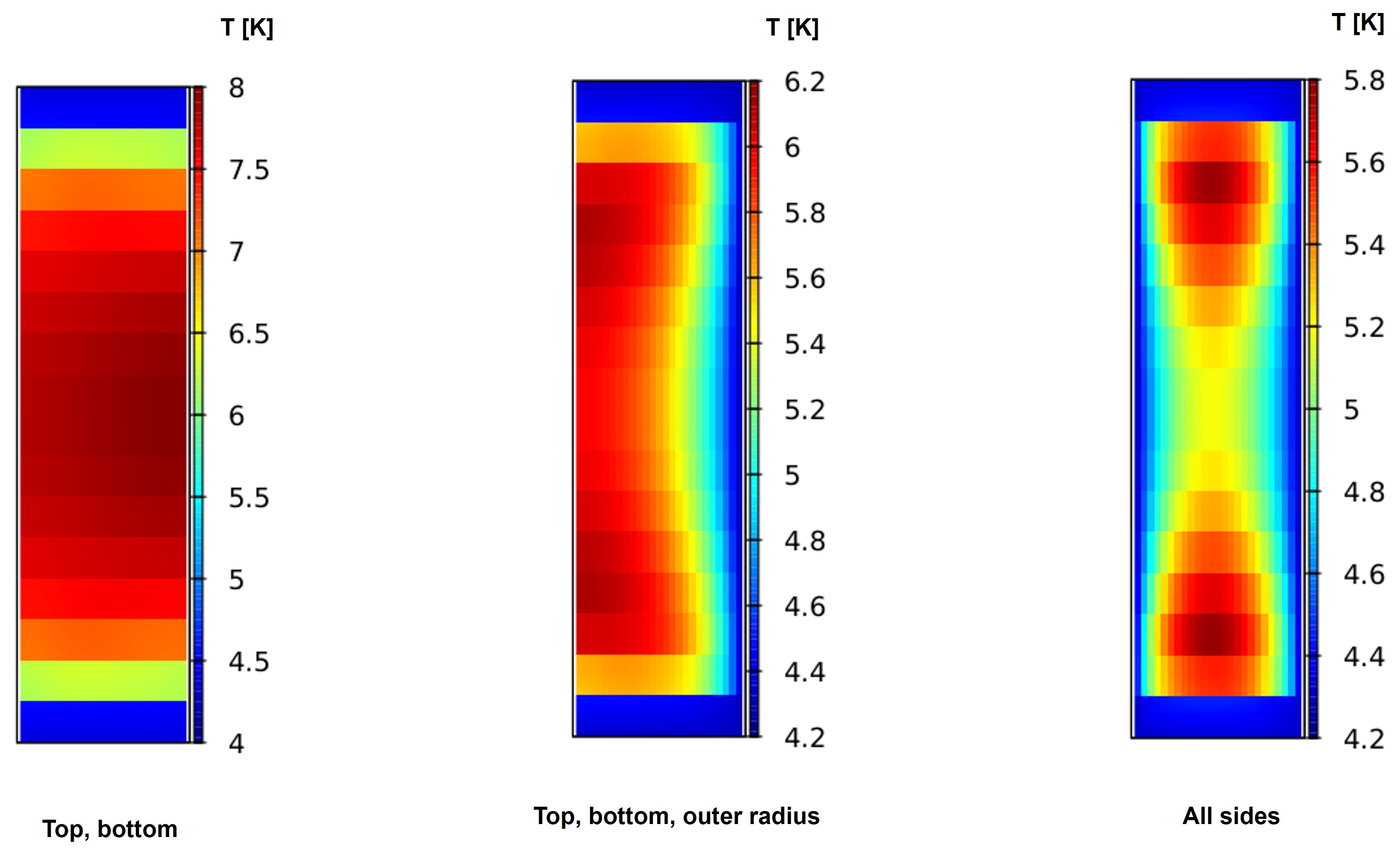}}

		\centering
	\subfloat[][]
	{\includegraphics[trim= 0 0 0 0, clip, width = 12 cm]{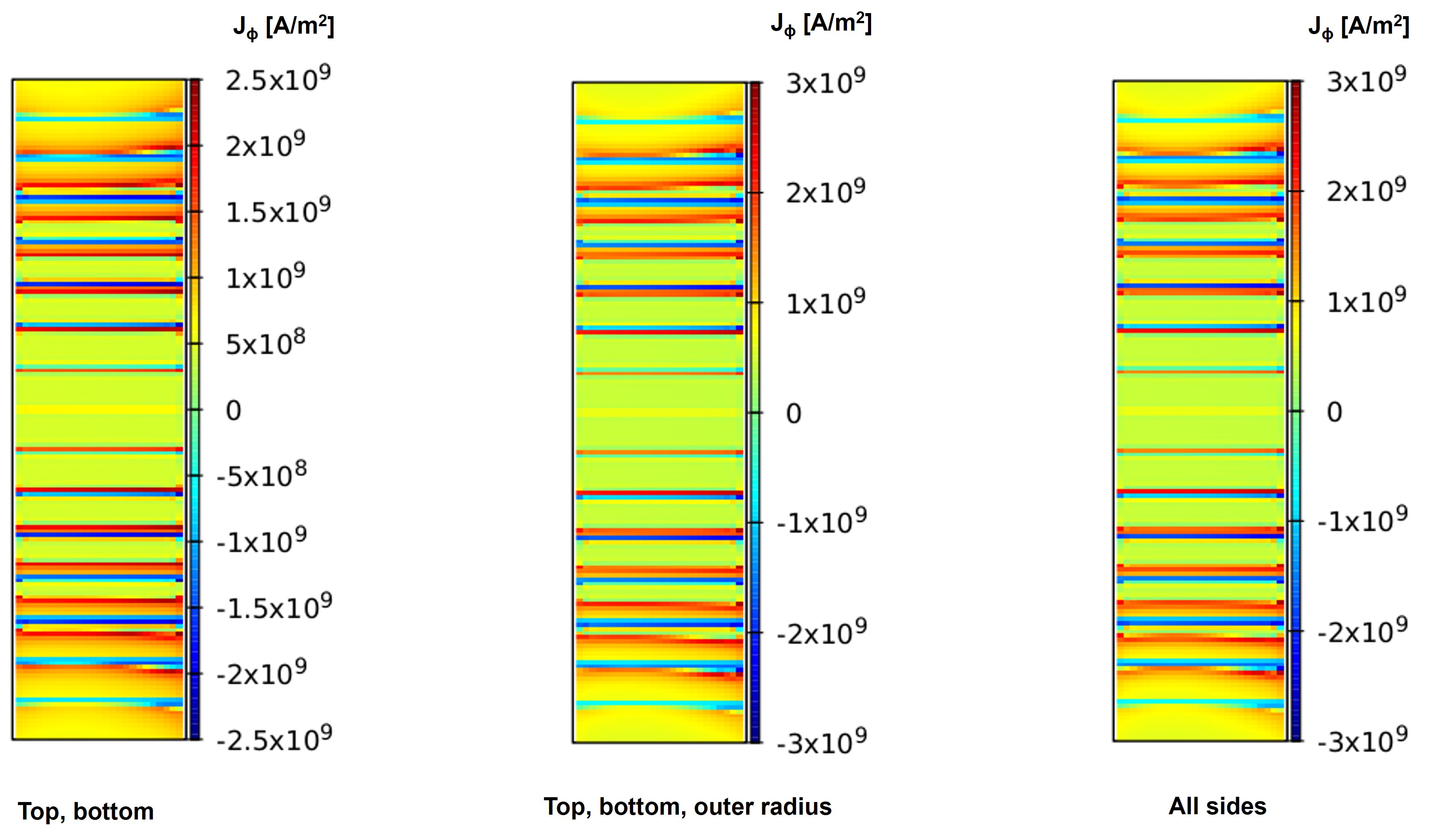}}

\caption{  (a) Total power and (b) Average temperature for several cooling scenarios of Figure \ref{hcurve} (b). (c) and (d) shows temperature and current density profiles at 333 s, where no quench is observed. } 
\label{cooling}
\end{figure*}

The results for different cooling boundary conditions (Figure \ref{hcurve} (b)(2-4)) are shown in Figure \ref{cooling}. Here we discuss both the temperature increase and the power, as they influence each other. With proper cooling boundary conditions, this rapid rise in temperature and power is controlled in other cases. As expected, the configuration with cooling on all sides presents the lowest temperature rise (see Figure \ref{cooling} (c) for the end of the ramp). The reason for this behaviour is the proper cooling of top and bottom pancakes in these cases (Figure \ref{cooling} (c)) where the temperature rise should be the highest, as seen in adiabatic case. As seen in Figure \ref{cooling} (a), the total power generated is much higher in the adiabatic case compard with the configurations with liquid helum cooling. This is due to the higher temperatures in the adiabatic case; which decreases the local $J_c$, and it increases the local electric field and the power density, ${\bf E}\cdot{\bf J}$. As a consequence of the higher power, it caues a higher temperature increase. With proper cooling at the boundaries, this rapid rise in temperature and power is under control. The current density profiles (Figure \ref{cooling} (d)) also show that the screening currents penetrate from the top and bottom pancakes, where the most power loss is concentrated. These are cooled now, as compared to the adiabatic case.  Due to that, the magnet is still superconducting at the end of the current ramp (Figure \ref{cooling} (d)), with little difference in the current density profiles between all scenarios with liquid helium cooling. {The uniform current density at the end pancakes is due to the loss in screening currents due to high power loss at the peak of current ramp.} At the plateau (above 333 s), there is no rise in current, and hence there is no AC loss except a residual contribution from decaying screening currents, and thus the temperature drops down to a stable plateau of 4.2 K (Figure \ref{cooling} (a) and (b)).

As seen from the temperature maps, the maximum temperature is between 5-8 K in these cases. When looking at the $h$ curve of liquid Helium, cooling is the most efficient in this range, which also contributes to stabilizing the temperature at these values. 

\subsection{Effect of damaged turns in metal-insulated coils} 

We also perform another study for several turn-to-turn resistances. This study is applied to a case where a turn (or group of turns) is damaged. In our case, we consider that there is a damaged homogenized turn at the bottom pancake, turn 5, which corresponds to actual turns 41-50. The reason for choosing the damaged turn at this location is that the temperature rise is the highest at the edge pancakes, as we saw before, and thus there is a higher possibility of damage at these pancakes. We assume that $J_c$ of the damaged turn is 10 percent of the original $J_c$. The different turn-to-turn contact resistances ($r$) are 10$^{-6}~\ohm \cdot \rm m^2$, 10$^{-7}~\ohm \cdot \rm m^2$ (metal insulated case), and 10$^{-8}~\ohm \cdot \rm m^2$ (non insulated case). Cooling is applied at top, bottom, and right edges of the magnet for this study. 

\begin{figure*} [tbp]

\begin{multicols}{2}    
		\centering
	\subfloat[][]
	{\includegraphics[trim= 0 0 0 0, clip, width = 8 cm]{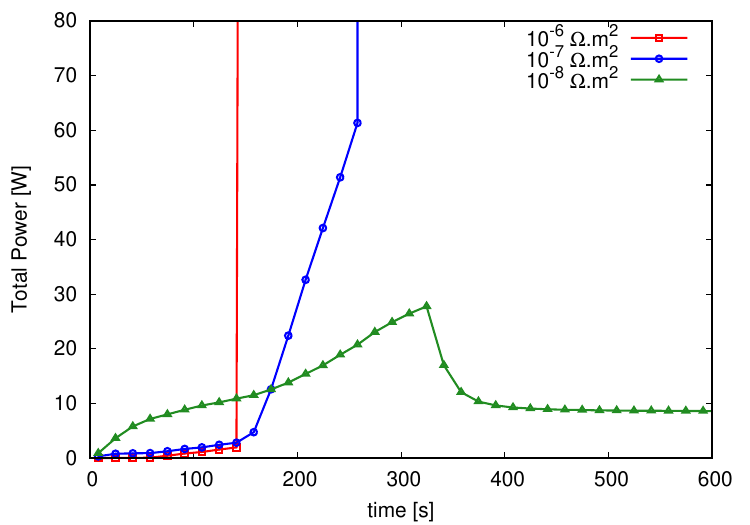}}
	\subfloat[][]
	{\includegraphics[trim= 0 0 0 0, clip, width = 8 cm]{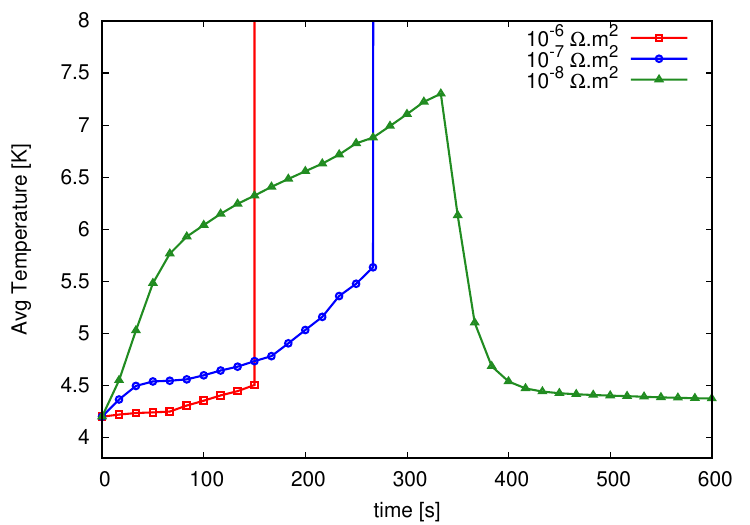}}	 
\end{multicols}

		\centering
	\subfloat[][]
	{\includegraphics[trim= 0 0 0 0, clip, width = 8 cm]{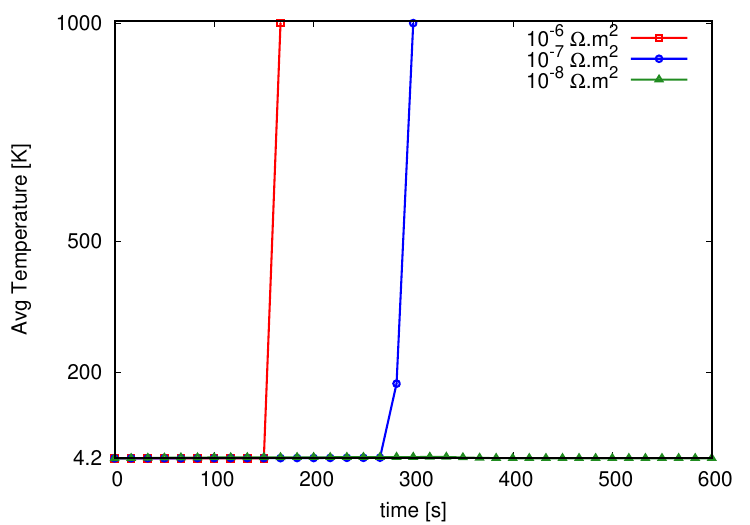}}

\begin{multicols}{2}    
		\centering
	\subfloat[][]
	{\includegraphics[trim= 0 0 0 0, clip, width = 8 cm]{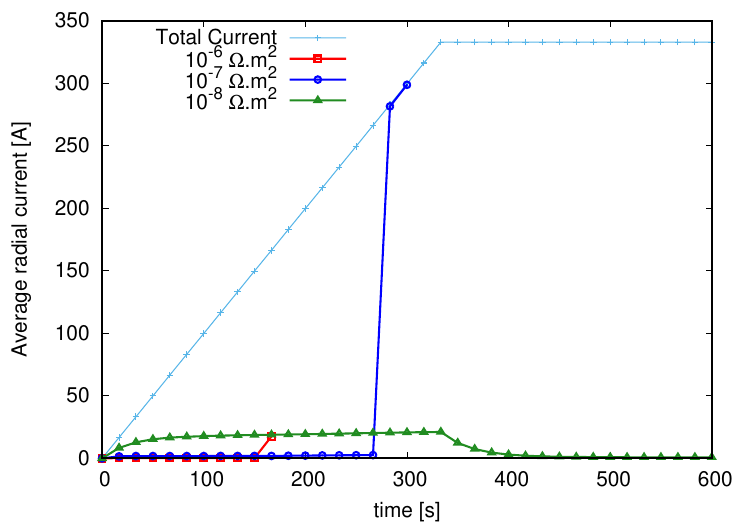}}
	\subfloat[][]
	{\includegraphics[trim= 0 0 0 0, clip, width = 8 cm]{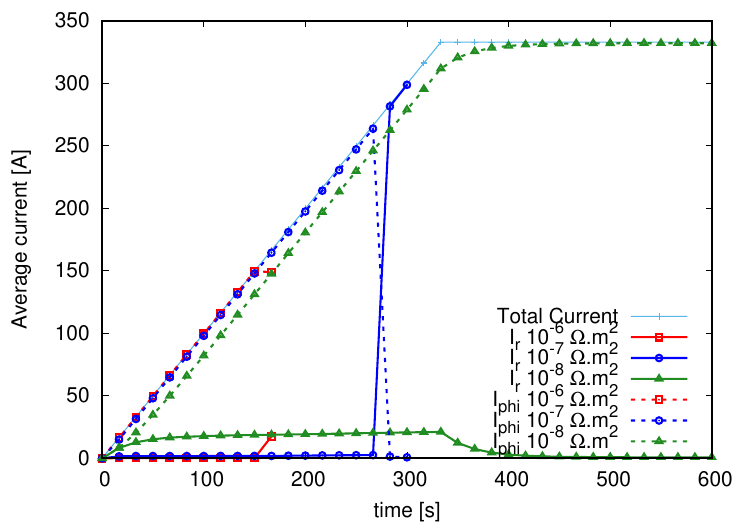}}	 
\end{multicols}

\caption{ Modelling results for damaged turns for (a) total power, and (b) average temperature for
different turn-to-turn contact resistances r. (c) shows the average temperature like in (b) but for higher temperature scale, showing that the maximum temperature reaches above 1000 K for high r. (d) and (e) shows the average radial, angular and total currents. } 
\label{dam1}
\end{figure*}

Firstly, the total power and average temperature can be seen in Figure \ref{dam1} (a), (b), and (c). The graphs show that the power generated is the lowest in the case of 10$^{-8}~\ohm \cdot \rm m^2$, which leads to the lowest temperature rise when compared to other turn-to-turn resistivities. For the other cases, there is a sharp rise in total power and average temperature, which leads to premature quench of the magnet (average temperature over 1000 K under 300 s), as seen in Figure \ref{dam1} (b) and (c). Thus, the possibility for quench reduces when the turn-to-turn resistivity is low, or it can be said that the non-insulated coils have lesser possibility of quench than the metal insulated coils. The reason is that, at a given temperature, when the current overcomes the critical current for a certain turn the local power dissipation increases with the contact resistance between turns \cite{pardoE2024SSTa}. Thus, heating increases with the contact resistance, and so does the temperature. Therefore, the minimum current to cause electrothermal quench decreases with the contact resistance (Figure \ref{dam1} (d)). After quench, there appears a sudden increase in radial current, due to the transition into high-resistive normal state of the superconductor (Figure \ref{dam1} (d)). As we impose current conservation in our model \cite{pardoE2024SSTa}, the rise in radial currents ($I_r$) results in drop in angular currents- $I_{phi}$ (Figure \ref{dam1} (e)). The calculations are stopped at the point of quench, where the temperature increases above 1000 K, and no futher points are plotted in the graph after that. 

\begin{figure*} [tbp]

		\centering
	{\includegraphics[trim= 0 0 0 0, clip, width = 7 cm]{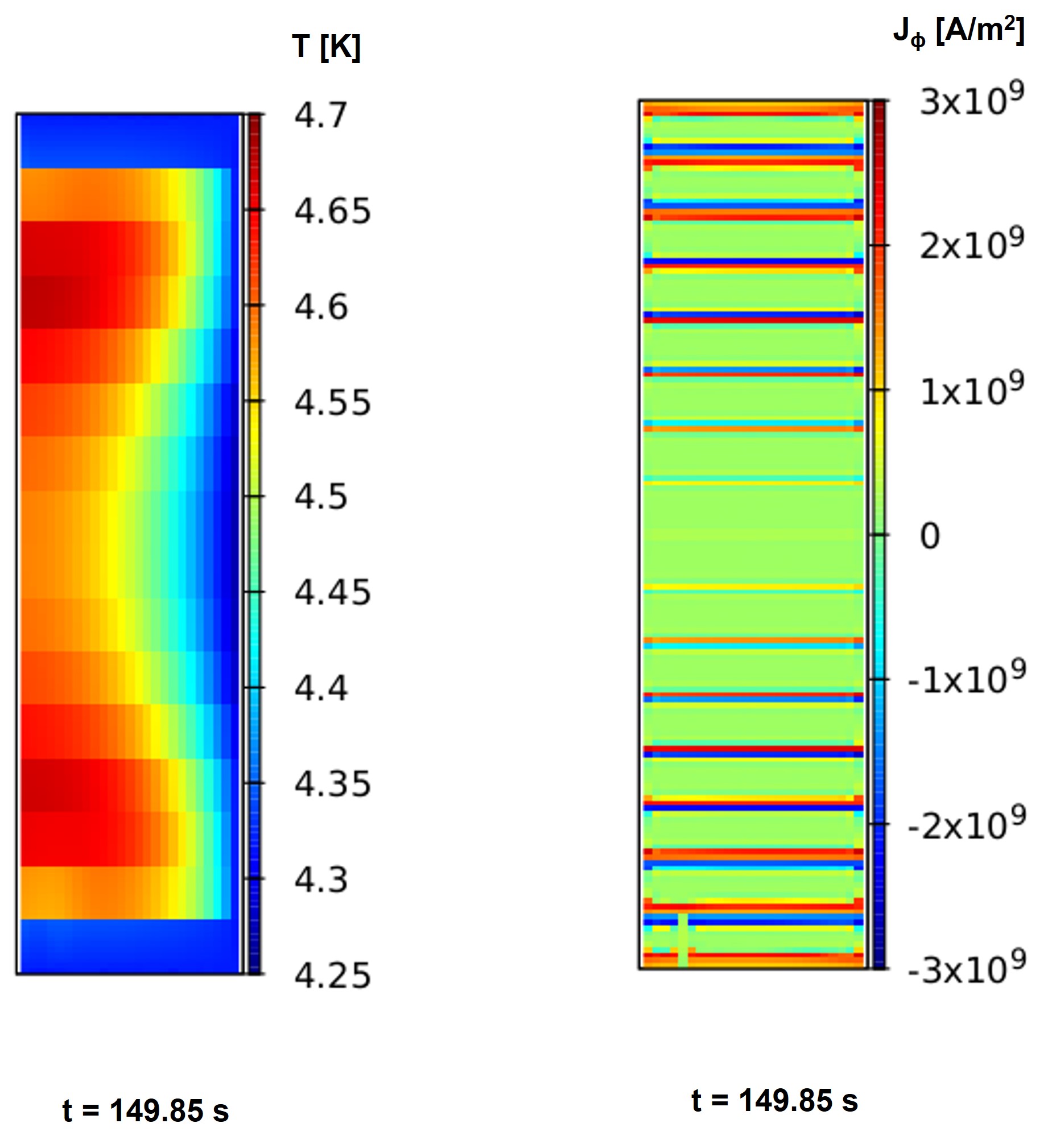}}

\caption{ Temperature and angular current density distributions for damaged turns (homogenized turn
5 at the bottom pancake) for contact resistance r=10$^{-6}~\ohm \cdot \rm m^2$ just before quench. } 
\label{dam2}
\end{figure*}

\begin{figure*} [tbp]

		\centering
	\subfloat[][]
	{\includegraphics[trim= 0 0 0 0, clip, width = 12 cm]{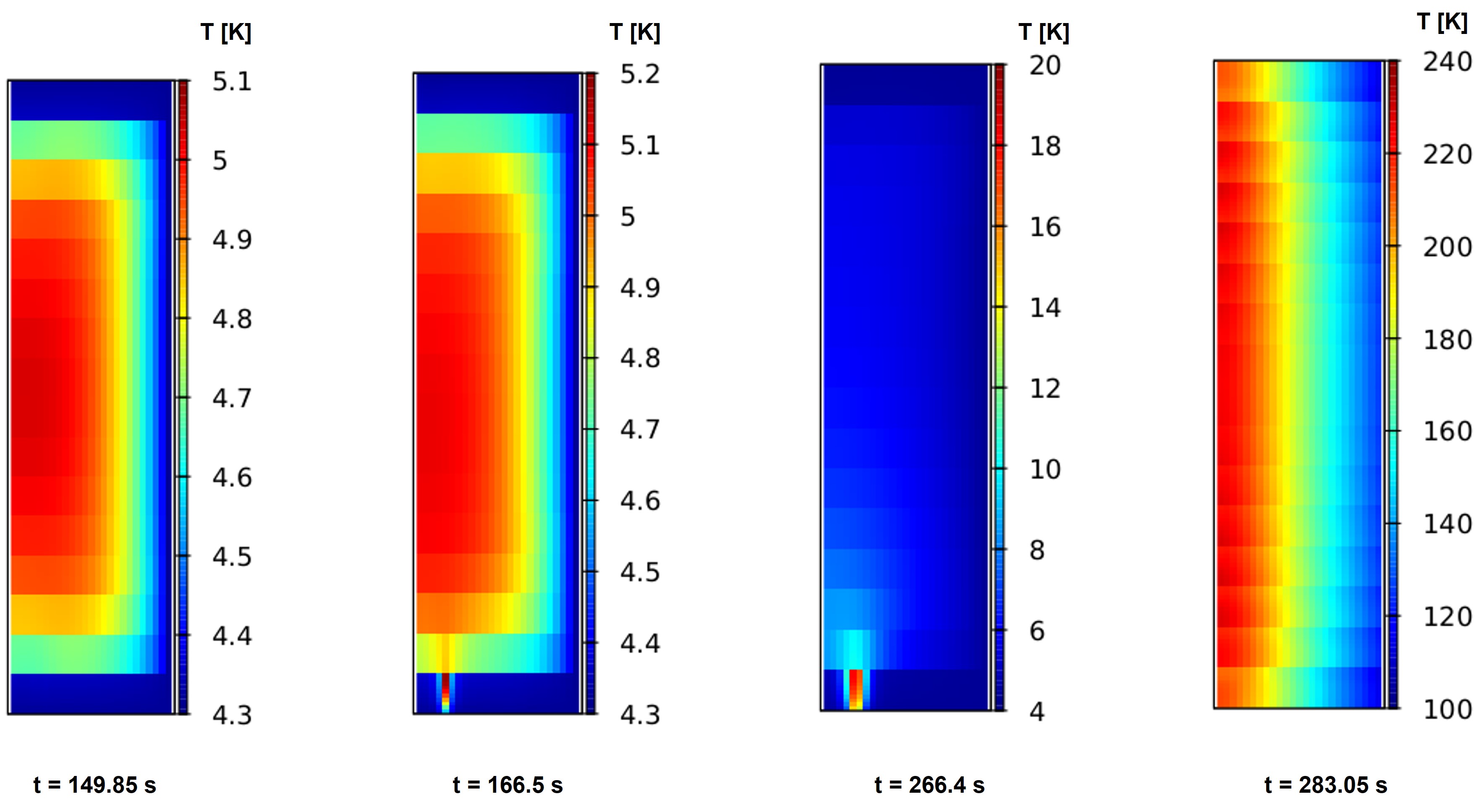}}

		\centering
	\subfloat[][]
	{\includegraphics[trim= 0 0 0 0, clip, width = 12 cm]{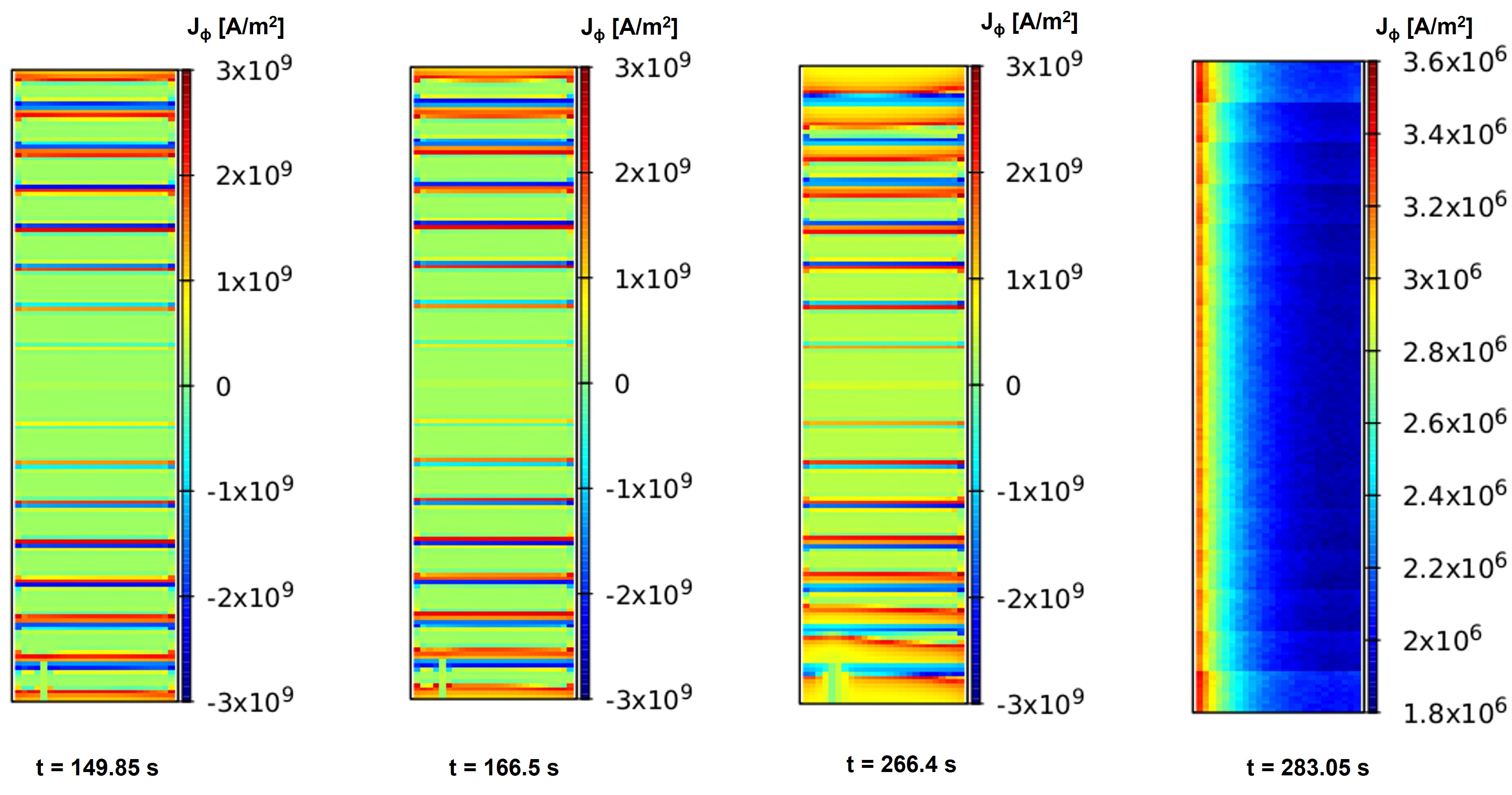}}

\caption{ Temperature (a) and angular current density (b) distributions for damaged turns
(homogenized turn 5 at the bottom pancake) for contact resistance r=10$^{-7}~\ohm \cdot \rm m^2$ just before and after quench. } 
\label{dam3}
\end{figure*}

\begin{figure*} [tbp]

		\centering
	\subfloat[][]
	{\includegraphics[trim= 0 0 0 0, clip, width = 12 cm]{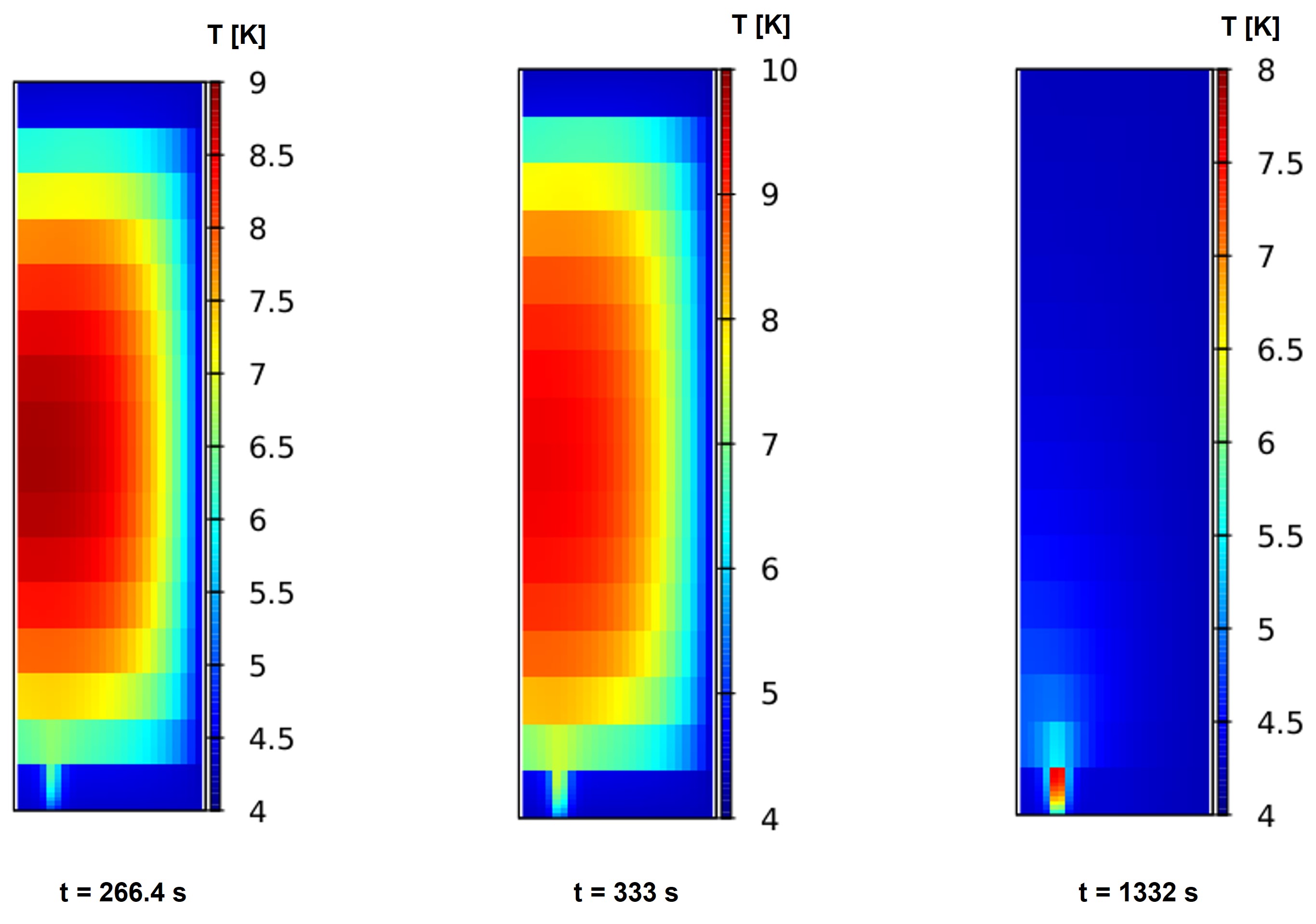}}

		\centering
	\subfloat[][]
	{\includegraphics[trim= 0 0 0 0, clip, width = 12 cm]{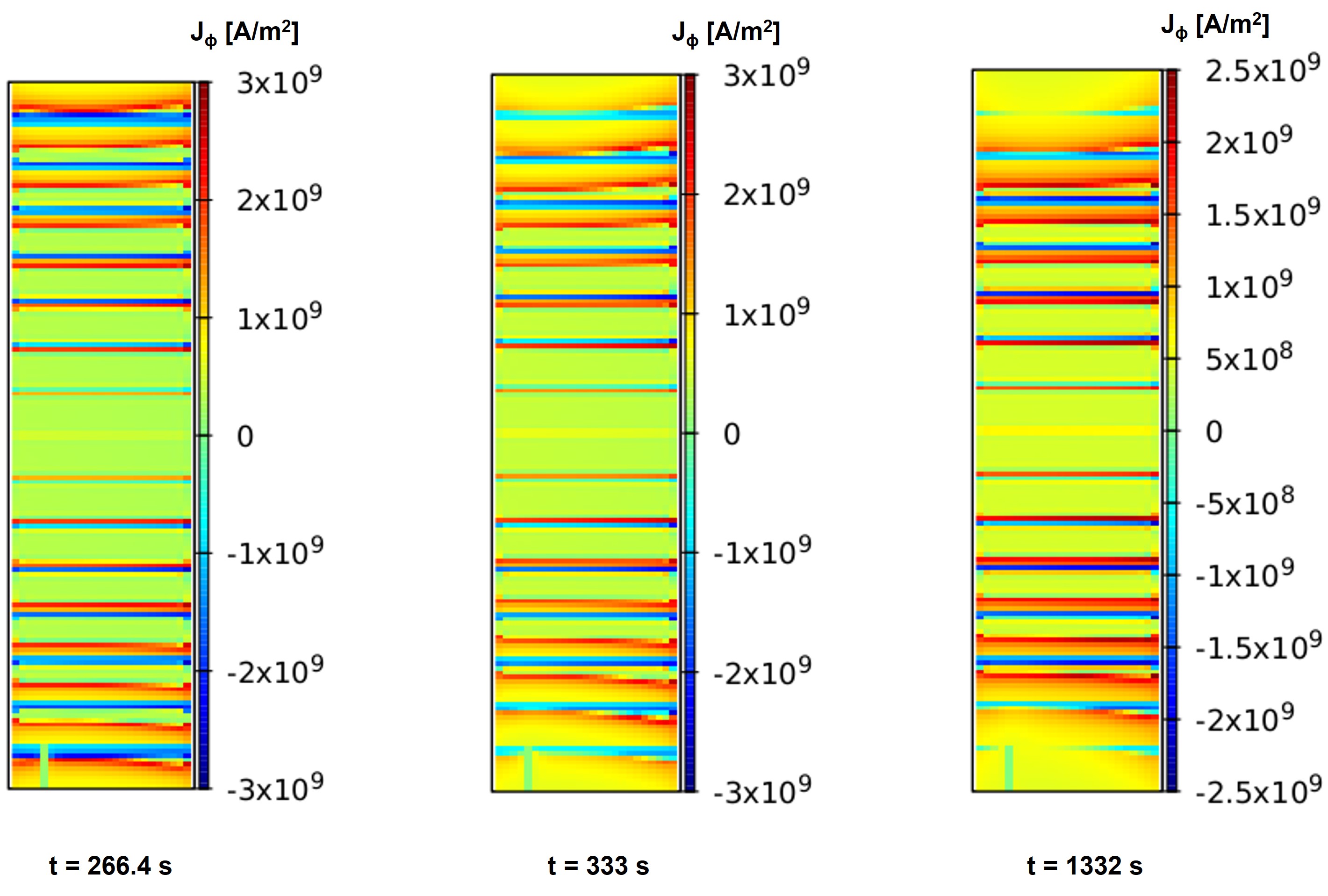}}

\caption{ The same as Figure \ref{dam3} but for contact resistance r=10$^{-8}~\ohm \cdot \rm m^2$, which is typical for non-insulated coils.  } 
\label{dam4}
\end{figure*}

\begin{figure*} [tbp]

\centering
	{\includegraphics[trim= 0 0 0 0, clip, width = 15 cm]{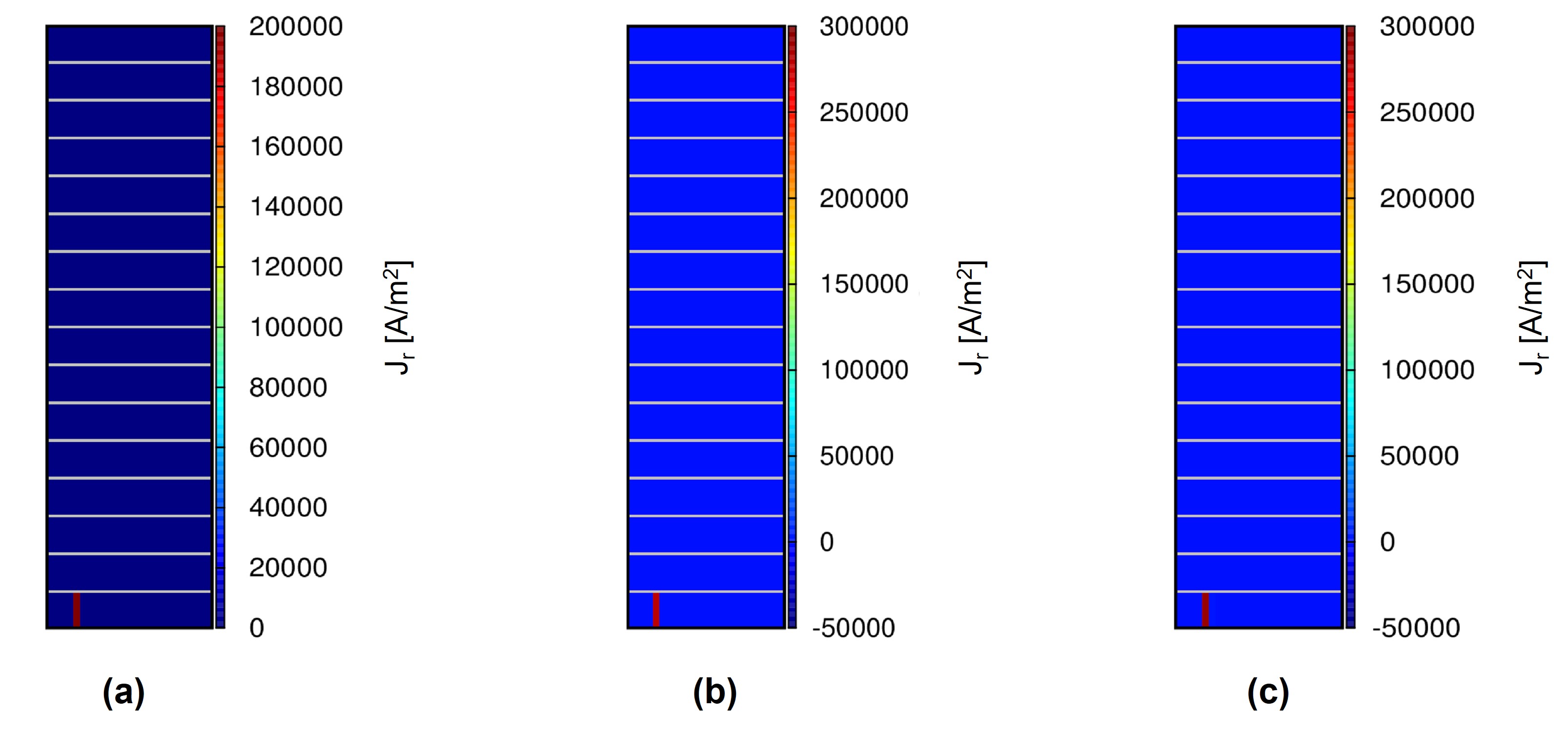}}

\caption{ Radial currents at the last calculated timestep for turn-to turn contact resistances (a) 10$^{-6}~\ohm \cdot \rm m^2$, (b) 10$^{-7}~\ohm \cdot \rm m^2$, and (c) 10$^{-8}~\ohm \cdot \rm m^2$. } 
\label{dam5}
\end{figure*}

The temperature and current density distribution for this study can be seen in Figures \ref{dam2}, \ref{dam3}, \ref{dam4}, and \ref{dam5}. For $r$ = 10$^{-6}~\ohm \cdot \rm m^2$, the magnet runs at stable temperatue only up to 149 s and quenches afterwards due to the damaged turn (Figure \ref{dam2}). For $r$ = 10$^{-7} ~\ohm \cdot \rm m^2$, the magnet can operate up to 266 s before quench (Figure \ref{dam3}). For this case, the maximum average temperature before quenching is at the damaged turn (around 20 K), where even cooling at the damaged turns on the magnet top and bottom edges is not effective enough to avoid thermal runaway. When looking at the current density, the radial currents are the highest in the damaged turns (Figure \ref{dam5}), which reduces the angular screening currents in the damaged turns (Figure \ref{dam3} and \ref{dam4}). These are the same regions where the temperature rise is the highest, as seen in these figures.

\R{Additionally, there is a very minimal temperature rise before quench for 10$^{-6} \ohm m^2$ , and later the magnet quenches very fast due to high radial currents and power generation (Fig. \ref{dam1}). The radial currents are higher for the configurations with lower turn-to-turn resistances, which gives significant rise to temperature already before quench. For the high contact resistance of 10$^{-6}$ $\Omega$m$^2$ there will also appear a temperature rise just before thermal quench, but this will happen much closer to quench than that for lower contact resistances. The high time stepping of the computations of this article is not able to catch this preliminary temperature rise of the pre-quench process.}

However, for non insulated coils ($r$ = 10$^{-8} ~\ohm \cdot \rm m^2$), the maximum temperature (around 7 K) is much lower than the metal insulated case, and hence it does not quench. The calculations can thus run for the whole operation of the magnet, including the current ramp plateau (till 1000 s and over), and the maximum average temperature is still not above 8 K at the damaged turns (Figure \ref{dam4}). This shows that the damaged turns in the magnet can be tolerated, as long as the turn-to-turn contact resistance is low enough. For our configuration, this occurs for the non-insulated coils.

\section{Conclusion}

A novel coupled electrothermal software is developed using variational principles and finite difference method. We see that appropriate boundary cooling conditions are essential for quench modeling due to low heat capacities at low temperatures. It is also shown that the non insulated coils behave thermally better than the metal insulated coils in the case of a damage in the magnet, due to low radial turn-to-turn resistances. Thus, the developed software can successfully simulate the quench behaviour of full scale HTS high field magnets, and it can be an effective tool for magnet designers performing electro-magneto-thermal quench simulations.

\section{Acknowledgements}

We acknowledge Oxford Instruments for providing details on the cross-section of the LTS outsert. We also acknowledge Mr. Lubomir Kopera for collecting data and developing graphs regarding liquid Helium $h$ curves. This project has received funding from the European Union's Horizon 2020 research and innovation programme under grant agreement No 951714 (superEMFL), and the Slovak Republic from projects APVV-19-0536 and VEGA 2/0098/24. Any dissemination of results reflects only the authors' view and the European Commission is not responsible for any use that may be made of the information it contains.

\end{document}